\documentclass[11pt]{article}
\usepackage{fullpage}

\pdfoutput=1
\usepackage{amssymb}
\usepackage{amsmath}
\usepackage{amsfonts}
\usepackage{xcolor} 
\usepackage{graphicx}
\usepackage{subfig}

\usepackage[T1]{fontenc}
\usepackage[utf8]{inputenc}
\usepackage[normalem]{ulem}
\usepackage{mathtools}

\usepackage{verbatim}
\usepackage{multirow}
\usepackage{dsfont}
\usepackage{cancel}
\usepackage{bm}
\usepackage{bbm}

\numberwithin{equation}{section}
\usepackage{slashed}
\usepackage{tabularx}
\usepackage{dcolumn}
\usepackage{hyperref}
\usepackage{setspace}
\usepackage{xspace}
\usepackage[bbgreekl]{mathbbol}
\usepackage{booktabs}
\usepackage{makecell}
\usepackage{array}

\usepackage{mathrsfs}
\usepackage{bbold}              
\usepackage{url}
\usepackage{float}
\usepackage{cite}

\newcommand{\eVdist}{\kern-0.06em}

\newcommand{\im}{\:\text{Im}\,}

\begin{document}

\begin{center}

{\Large \bf
Dirac Materials for Sub-MeV Dark Matter Detection: New Targets and Improved Formalism} \\
\vspace{0.5cm}

 { R.\ Matthias Geilhufe${}^{1}$, Felix Kahlhoefer${}^{2}$ and Martin Wolfgang Winkler${}^{3}$}

 \vspace*{.4cm}
${}^1$ {\it Nordita,  KTH Royal Institute of Technology and Stockholm University, Roslagstullsbacken 23,  10691 Stockholm,  Sweden} \\
${}^2${\it Institute for Theoretical Particle Physics and Cosmology (TTK), RWTH Aachen University, D-52056 Aachen, Germany} \\
${}^3$ {\it The Oskar Klein Centre for Cosmoparticle Physics, Department of Physics, Stockholm University, Alba Nova, 10691 Stockholm, Sweden}\\
\vspace*{.5cm}

\end{center}

\begin{abstract}\noindent

Because of their tiny band gaps Dirac materials promise to improve the sensitivity for dark matter particles in the sub-MeV mass range by many orders of magnitude. Here we study several candidate materials and calculate the expected rates for dark matter scattering via light and heavy dark photons as well as for dark photon absorption. A particular emphasis is placed on how to distinguish a dark matter signal from background by searching for the characteristic daily modulation of the signal, which arises from the directional sensitivity of anisotropic materials in combination with the rotation of the Earth. We revisit and improve previous calculations and propose two new candidate Dirac materials: BNQ-TTF and Yb$_3$PbO. We perform detailed calculations of the band structures of these materials and of ZrTe$_5$ based on density functional theory and determine the band gap, the Fermi velocities and the dielectric tensor. We show that in both ZrTe$_5$ and BNQ-TTF the amplitude of the daily modulation can be larger than 10\% of the total rate, allowing to probe the preferred regions of parameter space even in the presence of sizeable backgrounds. BNQ-TTF is found to be particularly sensitive to small dark matter masses (below $100\,\mathrm{keV}$ for scattering and below $50\,\mathrm{meV}$ for absorption), while Yb$_3$PbO performs best for heavier particles.

\end{abstract}

\vspace*{3mm}


\section{Introduction}

The realization that quantum materials, which have been the subject of great attention in recent years, may offer unique opportunities to search for light and very weakly interacting particles has led to a fruitful collaboration between particle physics and condensed matter physics. This development has given new hope to the ongoing search for dark matter (DM) in a time when experimental null results mount increasing pressure on traditional DM models (see e.g.~\cite{Athron:2018hpc}). Indeed, many novel detection strategies have been developed that promise to probe DM models in regions of parameter space that were previously thought to be experimentally inaccessible~\cite{Knapen:2017xzo,Lin:2019uvt}. This is especially true for DM particles with mass in the keV to MeV range, which would carry so little kinetic energy in the present Universe that their interactions with conventional detectors would be unobservable. While such particles are too light to be produced via the conventional freeze-out mechanism, recent studies have explored many alternative ways to reproduce the observed DM relic abundance, for example via the freeze-in mechanism~\cite{Hall:2009bx,Essig:2011nj,Chu:2011be,Green:2017ybv,Heeba:2018wtf,An:2018nvz,Dvorkin:2019zdi}. 

Given the typical velocity of DM particles in the solar neighborhood of $v = 10^{-3} c$, one needs to achieve an energy threshold of less than an eV in order to search for DM particles in the sub-MeV range. Among the proposed materials to achieve this goal are superconductors~\cite{Hochberg:2015pha,Hochberg:2016ajh,Hochberg:2019cyy}, superfluids~\cite{Schutz:2016tid,Knapen:2016cue,Caputo:2019cyg}, polar crystals~\cite{Knapen:2017ekk,Griffin:2018bjn,Cox:2019cod}, topological materials \cite{vergnioryDM}, and finally Dirac materials~\cite{hochberg,Coskuner:2019odd,geilhufe2018materials}, which are the topic of the present work. Dirac materials are defined as materials where the elementary excitations can be effectively described via the Dirac equation \cite{wehling2014dirac} with the relativistic flat-metric energy momentum relation
\begin{equation}
 E_\mathbf{k}^\pm = \pm \sqrt{v_\mathrm{F}^2 \, \mathbf{k}^2 + \Delta^2} \; ,
 \label{eq:dispersion}
\end{equation}
where $\mathbf{k}$ denotes the lattice momentum, $v_\mathrm{F}$ is the Fermi velocity (replacing the speed of light) and $2\Delta$ is the band gap (replacing the rest mass). For $|\mathbf{k}| \gg \Delta$ the electrons hence have a linear dispersion relation with coefficient of proportionality given by $v_\mathrm{F}$. 

A crucial advantage of Dirac materials is that the band gap $2\Delta$, which determines the energy threshold of the material, can be of the order of a few meV. Such small band gaps can arise for example in Dirac semimetals, when a spin degeneracy is lifted by weak spin-orbit coupling or if the underlying symmetry protecting the Dirac node is lifted. A band gap of this magnitude is ideal for the detection of sub-MeV DM particles while at the same time suppressing backgrounds from thermal excitations of electrons. Nevertheless, there are at present no realistic estimates of the expected background level in a Dirac material and existing sensitivity studies in the literature are based on the assumption that backgrounds can be neglected. 
This might be too optimistic since even in almost perfectly clean samples, states arising in tiny islands of impurity regions can lead to an exponentially small density of states in the mass gap of a Dirac semimetal~\cite{olsthoorn2019mass,dora2009gaps}. While this effect is usually negligible, it might play a significant role in rare event searches. As long as one is solely interested in deriving exclusion limits, it may still be justified to ignore backgrounds. However it arises
the question how the DM nature of a potential signal can be confirmed.

In the present work we explore how this question can be answered by searching for a daily modulation in the data. While such a modulation is absent for most backgrounds, it is expected for a DM signal because of the rotation of the Earth~\cite{Hochberg:2016ntt,Griffin:2018bjn,Coskuner:2019odd}. In combination with the motion of the Sun through the Milky Way this rotation leads to a ``DM wind'' in the laboratory frame that changes its direction over the course of each day. Provided the detector is anisotropic, i.e.\ that its response depends on the direction of the momentum transfer $\mathbf{q}$, the resulting modulation may allow to confirm the DM origin of an observed signal. 

In Dirac materials such an anisotropy arises from the fact that both the Fermi velocities and the dielectric constants typically differ for the different directions in reciprocal space. It was shown in Ref.~\cite{hochberg} that as a result scattering in certain directions may be heavily suppressed or even kinematically forbidden, which makes these materials ideally suited to search for daily modulations. In this work we develop the necessary formalism to calculate the modulation of the DM signal and point out a number of subtleties overlooked in previous studies. We furthermore identify the regions of parameter space of specific models of DM where the modulation is large enough to be detected with statistical significance.

Throughout the paper we will discuss three Dirac materials as potential sensor materials for DM detection. First, ZrTe$_5$, which was initially discussed in connection to Dirac materials for DM sensors due to its tiny and well isolated direct gap \cite{hochberg}. Second, we consider the $f$-electron antiperovskite Yb$_3$PbO which was found to exhibit massive Dirac states along certain high symmetry paths in the Brillouin zone \cite{pertsova2019}. Third we follow the outcome of the materials informatics approach to identify potential dark matter sensor materials discussed in Ref. \cite{geilhufe2018materials} and reveal that one of the three materials mentioned in the study, the quasi 2-dimensional organic molecular crystal BNQ-TTF, exhibits various Dirac crossings within the Brillouin zone when spin-orbit coupling is taken into account. These nodes can potentially be gaped by applying stress and as a result breaking some of the crystalline symmetries protecting the Dirac nodes.

In addition to the scattering of sub-MeV DM particles, we also discuss the absorption of bosonic relics with sub-eV masses. We point out that -- in contrast to previous claims -- the modulation of the signal is absent in this case.

This work is structured as follows. In Sec.~\ref{sec:formalism} we present the general formalism for the calculation of the expected event rate and its daily modulation, both for the case of DM scattering and absorption. Sec.~\ref{sec:polarizationtensor} provides an improved calculation of the polarization tensor for anisotropic Dirac materials. Our numerical calculations of the properties of several candidate Dirac materials are discussed in Sec.~\ref{sec:candidates}. In Sec.~\ref{sec:results} we then introduce the statistical method that we employ and present our sensitivity estimates. Finally, we summarize our findings in Sec.~\ref{sec:summary}.

\section{Dark Matter Interactions in Dirac Materials}
\label{sec:formalism}

While Dirac materials can in principle be used to probe many different models of sub-MeV DM, they are particularly well suited for probing $U(1)$ gauge extensions of the Standard Model. These extensions contain a dark photon $A^\prime$ which kinetically mixes with the ordinary photon via $\mathcal{L}\supset -\frac{\varepsilon}{2} F_{\mu\nu}F^{\prime\mu\nu}$, where $F_{\mu\nu}$ ($F'_{\mu\nu}$) denotes the field strength of the (dark) photon. The dark photon can either be a DM candidate itself or it can mediate the interactions between another DM particle and visible matter. The formalism to calculate the resulting detector signals for Dirac materials has been developed in~\cite{hochberg,Coskuner:2019odd}. For the case of anisotropic Dirac materials, however, we find a number of pertinent differences with the expressions provided in these works. We will therefore revisit the derivation of the event rates for DM scattering and absorption in detail and provide improved formulas.

\subsection{Scattering Rates in Dirac Materials}

We first consider a DM particle $\chi$ with mass $m_\chi$ which is charged under the new $U(1)$ gauge group. The total DM-electron scattering rate in a Dirac material with volume $V$ is given by
\begin{equation}
 R_{\text{tot}}= g\, V \,V_{\text{uc}} \,n_e \int\frac{\mathrm{d}^3 k \,\mathrm{d}^3 k^\prime}{(2\pi)^6} R_{\mathbf{k}\rightarrow\mathbf{k}^\prime}\,,
\end{equation}
where $n_e$ stands for the number of valence band electrons per unit mass and $V_{\text{uc}}$ for the volume of the unit cell. The factor $g=g_s\,g_C$ is the product of spin degeneracy $g_s$ and Dirac cone degeneracy $g_C$~\cite{hofmann2015,throckmorton}. The rate for lifting one electron with initial and final lattice momentum $\mathbf{k}$ and $\mathbf{k}^\prime$ from the valence band into the conduction band reads~\cite{Essig:2015cda}
\begin{align}\label{eq:Rkk}
 R_{\mathbf{k}\rightarrow\mathbf{k}^\prime} = \frac{\rho_\chi}{m_\chi}\frac{\bar{\sigma}_e}{8\pi\mu_{\chi e}^2}\int d^3q\:|F_{\text{DM}}(q)|^2|\mathcal{F}_{\text{med}}(q)|^2 |f_{\mathbf{k}\rightarrow\mathbf{k}^\prime}(q)|^2\: \frac{\tilde{g}\!\left(v_{\text{min}},\psi\right)}{|\mathbf{q}|} \,,
\end{align}
with the four-momentum transfer $q^\mu=(\omega,\mathbf{q})$ and $\mathbf{q} = \mathbf{k}' - \mathbf{k}$. The DM density is denoted by $\rho_\chi$ and the reduced mass of the DM-electron system by $\mu_{\chi e}$. Furthermore, the fiducial DM-electron cross section is defined as
\begin{equation}
 \bar{\sigma}_e = \frac{\mu_{\chi e}^2}{16\pi m_\chi^2 m_e^2}\,|\mathcal{M}_0(q_0)|^2\,.
\end{equation}
It is convenient to evaluate the matrix element $\mathcal{M}_0$ for scattering on a free electron at $q_0^2=\alpha^2\,m_e^2$, where $\alpha$ and $m_e$ stand for the fine structure constant and the electron mass respectively. The momentum-dependence of the scattering, which results from the propagator of the exchanged dark photon, is then pulled into the form factor~\cite{Essig:2011nj}
\begin{equation}
  F_{\text{DM}}(q)=\frac{\mathcal{M}_0(q)}{\mathcal{M}_0(q_0)}=\frac{q_0^2-m_{A^\prime}^2}{q^2-m_{A^\prime}^2}\,.
\end{equation} 
In the main part of this work, we will focus on the case of a very light dark photon and, therefore, neglect $m_{A^\prime}$ in this expression. The case of a heavy dark photon mediator will be covered in App.~\ref{sec:heavymediator}.

Next, we turn to the form factor $\mathcal{F}_{\text{med}}(q)$ which accounts for the optical response of the medium. More specifically, it parametrizes the ratio of the in-medium scattering amplitude $\mathcal{M}$ over the free amplitude
\begin{equation}
 \mathcal{F}_{\text{med}}(q)=\frac{\mathcal{M}(q)}{\mathcal{M}_0(q)}=\frac{j_\mu^\prime \:D^{\mu\nu}\: j_\nu}{j_\mu^\prime\: {D_0}^{\mu\nu} \:j_\nu}\simeq \frac{D^{00}}{{D_0}^{00}}\,,
\end{equation}
where $j^\prime$ and $j$ denote the DM and the electron current respectively. The difference compared to the vacuum case manifests in the appearance of the in-medium photon propagator $D$ instead of the free propagator $D_0$. In the last step, we used the fact that the scattering process is non-relativistic, which implies $j^0\gg|\mathbf{j}|$.\footnote{See derivation of the Coulomb potential in standard text books (e.g.~\cite{Peskin:1995ev})}
The in-medium photon propagator can be derived from the Schwinger-Dyson equation for the electromagnetic field~\cite{Dyson:1949ha,Schwinger:1951ex}
\begin{equation}\label{eq:Schwinger}
  D^{-1} = D_0^{-1} - i\Pi\,,
\end{equation}
where $\Pi$ stands for the photon polarization tensor. We will explicitly calculate $\Pi$ for Dirac materials in Sec.~\ref{sec:polarizationtensor}. As we will prove there, the spatial components of $\Pi$ are negligible in the kinematic regime $|\mathbf{q}|\gg\omega$ relevant for DM scattering. Therefore, we obtain
\begin{equation}\label{eq:propagatorscatter}
D^{00}(q)\simeq \frac{-i}{q^2-\Pi^{00}(q)}\;\;\Longrightarrow\;\; \mathcal{F}_{\text{med}}(q)= \frac{q^2}{q^2-\Pi^{00}(q)}\,.
\end{equation}
The scattering rate, furthermore, depends on the transition form factor $f_{\mathbf{k}\rightarrow\mathbf{k}^\prime}$, which results from the electron wave functions in the Dirac material~\cite{Essig:2015cda,hochberg},
\begin{equation}
 |f_{\mathbf{k}\rightarrow\mathbf{k}^\prime}(q)|^2 =
 \frac{(2\pi)^3}{2\,V} \left( 1-\frac{\tilde{\mathbf{k}} \tilde{\mathbf{k}}^\prime+\Delta^2}{\sqrt{\tilde{\mathbf{k}}^2+\Delta^2} \sqrt{\tilde{\mathbf{k}}^{\prime\: 2}+\Delta^2}}\right) \delta(\mathbf{q}-(\mathbf{k}-\mathbf{k}^\prime))\,,
\end{equation}
where $2\Delta$ is the energy gap between the valence band and the conduction band. The tilde indicates that each three-momentum component is rescaled with the Fermi velocity in the corresponding direction, for example $\tilde{\mathbf{k}}=(k_x\,v_{\mathrm{F},x},k_y\,v_{\mathrm{F},y},k_z\,v_{\mathrm{F},z})$. 

The last ingredient in eq.~\eqref{eq:Rkk} is the velocity integral, which arises from an integration over the DM velocity distribution $f(\mathbf{v})$:
\begin{equation}
\tilde{g}= 2|\mathbf{q}|\int f(\mathbf{v}
) \,\delta\left(E_\mathrm{f} - E_\mathrm{i}\right) \mathrm{d}^3 v  \; ,
\end{equation}
where the factor $2|\mathbf{q}|$ has been introduced for convenience. The total energy of the initial and final state are denoted by $E_\mathrm{i}$ and $E_\mathrm{f}$.
In the so-called Standard Halo Model, the DM velocity distribution is given by
\begin{equation}
 f(\mathbf{v}
 ) = N \exp \left( - \frac{(\mathbf{v} - \mathbf{v}_\mathrm{e})^2}{v_0^2} \right) \Theta\left(v_\text{esc} - |\mathbf{v} - \mathbf{v}_\mathrm{e}|\right) \; ,
\end{equation}
where $N$ is a normalization factor, $\mathbf{v}_\mathrm{e}$ is the Earth's velocity, $v_0$ and $v_\text{esc}$ are the velocity dispersion and the Galactic escape velocity and $\Theta$ denotes the Heaviside step function. Note that the velocity distribution only depends on $v = |\mathbf{v}|$ and $\cos \theta_e = \hat{\mathbf{v}}\cdot\hat{\mathbf{v}}_\mathrm{e} $ (the hat indicates unit vectors), i.e.\  $f(\mathbf{v}) =  f(v, \cos \theta_e)$.

In the non-relativistic limit, the initial and final energy are given by
\begin{align}
 E_\mathrm{i} & = m_\chi + m_e + \frac{1}{2} m_\chi v^2 - E_{\mathbf{k}}\,, \\
 E_\mathrm{f} & = m_\chi + m_e + \frac{(m_\chi \mathbf{v} - \mathbf{q})^2}{2 \, m_\chi} + E_{\mathbf{k}+\mathbf{q}}\,.
\end{align}
with $\mathbf{q}$ again denoting the momentum transfer and
\begin{equation}
  E_{\mathbf{k}} = \sqrt{ \tilde{\mathbf{k}}^2+\Delta^2}\,.
\end{equation}
We then find
\begin{equation}
 E_\mathrm{f} - E_\mathrm{i}  = E_{{\mathbf{k}}+\mathbf{q}} + E_{\mathbf{k}} + \frac{q^2}{2 \, m_\chi} - \mathbf{q} \cdot \mathbf{v}  \equiv |\mathbf{q}| \left(v_\text{min} - v \, \cos \theta_q \right) \; ,
\end{equation}
where we have introduced $\cos \theta_q = \hat{\mathbf{v}}\cdot\hat{\mathbf{q}}$ and the minimal velocity
\begin{equation}
 v_\text{min} = \frac{E_{{\boldsymbol k}+\mathbf{q}} + E_{\boldsymbol k}}{|\mathbf{q}|} + \frac{|\mathbf{q}|}{2 \, m_\chi} \; .
\end{equation}
The velocity integral can hence be written as
\begin{align}
\tilde{g} & = 2 \int \frac{\mathrm{d}^3v}{v} f(v, \cos \theta_e) \delta \left(\frac{v_\text{min}}{v} - \cos \theta_q \right)  \; .
\end{align}

\begin{figure}[t]
\begin{center}   
 \includegraphics[width=0.5\textwidth]{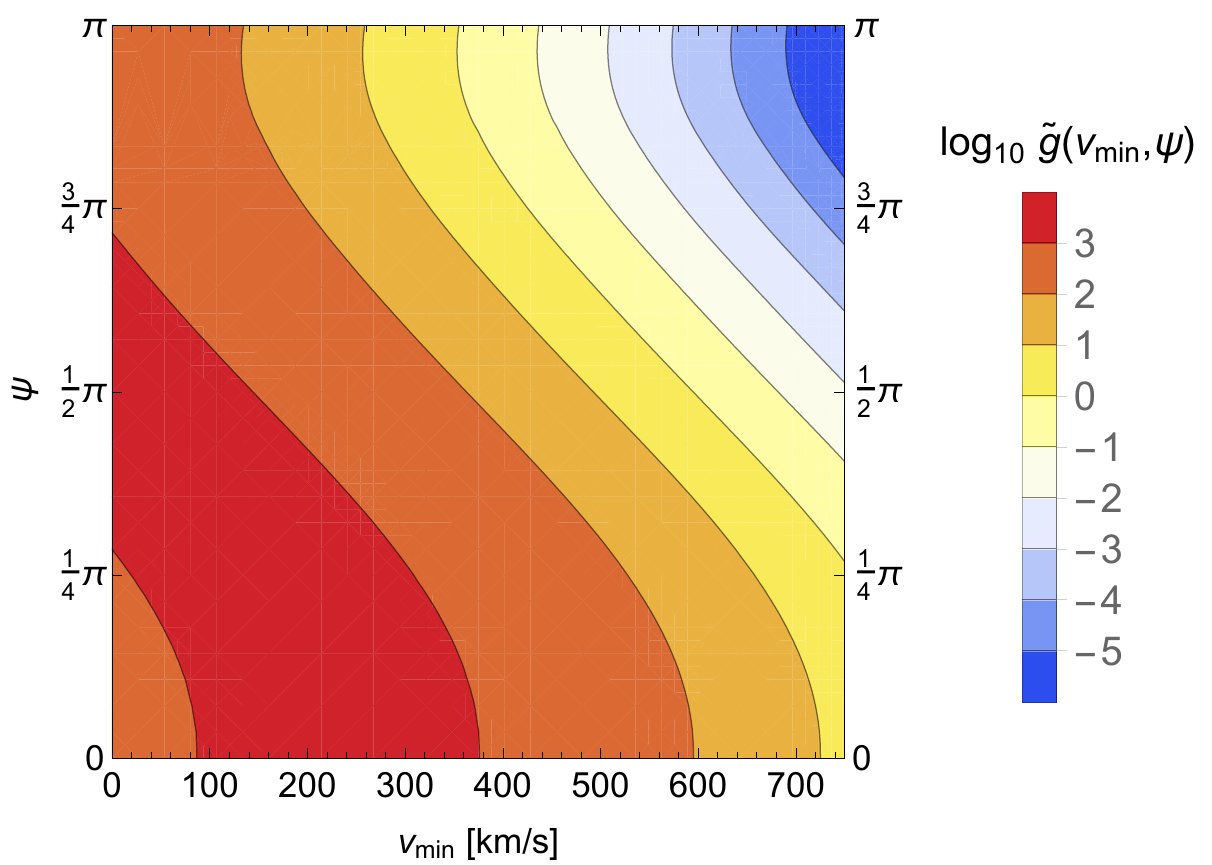}
\end{center}
\caption{The anisotropic velocity integral $\tilde{g}(v_\text{min},\psi)$.}
\label{fig:gtilde}
\end{figure}

Without loss of generality, we can choose the coordinate system such that the $z$-axis is aligned with $\mathbf{q}$. Furthermore, we require the earth velocity vector to reside in the $y$-$z$-plane. In spherical coordinates $(v, \theta, \phi)$ one then finds $\theta_q = \theta$ and 
\begin{equation}
\cos \theta_e = \sin \theta \, \sin \phi \, \sin \psi + \cos \theta \, \cos \psi \; ,
\end{equation}
where $\psi$ denotes the angle between $\mathbf{q}$ and $\mathbf{v}_\mathrm{e}$. The integration over $\cos \theta$ then yields 0 if $v < v_\text{min}$ and otherwise sets $\cos \theta = v_\text{min} / v$. We therefore find
\begin{equation}
 \tilde{g}=\tilde{g}(v_\text{min}, \psi) = \int_{v>v_\text{min}} 2\,v f\left(v, \sin \phi \, \sin \psi \sqrt{1 - \tfrac{v_\text{min}^2}{v^2}} + \cos \psi \frac{v_\text{min}}{v}\right)  \mathrm{d} v \mathrm{d} \phi \; .
 \label{eq:gtilde}
\end{equation}
An important feature of this result is that it does not depend on $|\mathbf{q}|$. Indeed $\tilde{g}$ is entirely determined by the two variables $v_\text{min}$ and $\psi$.
Calculating this integral numerically and tabulating the results as a function of two variables is straight-forward. The result is shown in Fig.~\ref{fig:gtilde} and confirms the naive expectation that scattering in the direction of the DM wind (i.e.\ $\psi \approx 0$) is strongly preferred.

We can now transform into the laboratory frame, in which $\mathbf{v}_\mathrm{e}$ is time-dependent. We adopt the same coordinate system as proposed in Ref.~\cite{Griffin:2018bjn}, in which $\mathbf{v}_\mathrm{e}$ points in the $z$-direction at $t = 0 \, \mathrm{days}$ and lies in the $y$-$z$-plane at $t = 0.5 \, \mathrm{days}$:
\begin{equation}
 \mathbf{v}_\mathrm{e}(t) = v_e \left( \begin{array}{c} \sin \alpha_e \sin \beta \\
 \sin \alpha_e \cos \alpha_e (\cos \beta - 1) \\
 \cos \alpha_e^2 + \sin \alpha_e^2 \cos \beta \end{array} \right)\;,
\end{equation}
where $\alpha_e = 42^\circ$ is the angle between the Earth's rotation axis and its velocity and $\beta = 2\pi \times t / 1 \mathrm{days}$. Finally, $\psi$ is obtained from $\cos\psi=\hat{\mathbf{v}}_\mathrm{e} \cdot \hat{\mathbf{q}}$.

In order to understand the impact of the Earth's rotation on the DM scattering rate, it is instructive to consider collisions with $\psi=0$ which dominate the velocity integral. For those, we can derive the inequality
\begin{equation}\label{eq:vmininequality}
    v_\text{min} \geq \sqrt{v_{\mathrm{F},i}^2+4\,\frac{\Delta}{m_\chi}}\,,
\end{equation}
with approximately $i=z$ ($i=y$) at $t=0$ ($t=0.5$). The fraction of DM particles which can undergo scattering, hence, strongly depends on the Fermi velocity in the direction of the DM wind. This implies strong daily modulations of the scattering rate in anisotropic Dirac materials with $v_{\mathrm{F},y}\neq v_{\mathrm{F},z}$.

A final subtlety arises from the fact that the analogy between the electron and a free Dirac fermion only applies for sufficiently small momenta $\mathbf{k}$. For larger momenta, the dispersion relation of the electron will deviate from eq.~\ref{eq:dispersion}. Of course, electrons with such large momenta may still contribute to the event rate, but the formalism outlined above cannot be applied. To obtain a conservative estimate of the event rate, Ref.~\cite{hochberg} introduced a cut-off $\Lambda$ and considered only scattering processes for which $\mathbf{k}, \mathbf{k'} < \Lambda$. For a known band structure the cut-off $\Lambda$ can be determined by identifying the momentum for which the dispersion relation becomes non-linear.

In the case of an anisotropic Dirac material the definition of $\Lambda$ becomes more subtle. Indeed, in this case the cut-off momentum typically depends on the direction, $\mathbf{\Lambda} = (\Lambda_x, \Lambda_y, \Lambda_z)$. While in principle it would be possible to apply different cut-offs in different directions, we will again adopt a simpler and more conservative approach and require
\begin{equation}
\tilde{k}, \tilde{k'} < \text{min}(\Lambda_x v_{\mathrm{F},x},\Lambda_y v_{\mathrm{F},y},\Lambda_z v_{\mathrm{F},z}) \equiv \tilde{\Lambda} \; . 
\label{eq:Lambdatilde}
\end{equation}
Note that this prescription differs from the one proposed in Ref.~\cite{hochberg}, where the maximum is taken rather than the minimum (presumably because of a typographical mistake).

\subsection{Absorption of Dark Photon Dark Matter}

Let us now consider the case that the dark photon itself constitutes the DM. It can then be absorbed in a Dirac material in analogy to the photoelectric effect. Specifically, we are interested in the absorption of non-relativistic dark photons with rest mass comparable to the band gap, which implies that $\omega\simeq m_{A^\prime} \gg |\mathbf{q}|$. In this regime -- as we will show in the next section -- the spatial components of the in-medium photon propagator can be approximated as
\begin{equation}\label{eq:propagatorabsorb}
  D^{ij} = \frac{-i g^{ij}}{q^2+\Pi^{ii}(q)}\,.
\end{equation}

We now want to determine the effective in-medium mixing angle $\varepsilon_{\text{med}}$ between the dark and the ordinary photon. For anisotropic materials $\varepsilon_{\text{med}}$ depends on the polarization. Since the dark photons are non-relativistic, we can conveniently choose the polarization vectors
\begin{equation}
 (\epsilon_x)^{\mu} = (0,1,0,0)\,,\qquad (\epsilon_y)^{\mu} = (0,0,1,0)\,,\qquad (\epsilon_z)^{\mu} = (0,0,0,1)\,.
\end{equation}
The in-medium mixing angle for an $x$-polarized dark photon is then obtained from the relation\footnote{The analogous expression for longitudinal, transverse polarization can e.g.\ be found in~\cite{An:2013yfc}.}
\begin{equation}
 \varepsilon_{\text{med},x}\, (\epsilon_{x})^\mu \equiv \varepsilon \, m_{A^\prime}^2\, D^{\mu\nu} \,(\epsilon_{x})_\nu\,,
\end{equation}
which implies
\begin{equation}
 |\varepsilon_{\text{med},x}|^2 = \varepsilon^2\,\frac{m_{A^\prime}^4}{\left|m_{A^\prime}^2+\Pi^{11}(m_{A^\prime})\right|^2 }\,.
\end{equation}
In the above expression we explicitly indicate that the polarization tensor has to be evaluated at $q^2=m_{A^\prime}^2$. The $x$-polarized dark photon absorption rate is determined from the optical theorem (see e.g.~\cite{An:2013yua})
\begin{equation}
\Gamma_x = \frac{|\varepsilon_{\text{med},x}|^2 \,(\epsilon_x)_\mu \im \Pi^{\mu\nu}\, (\epsilon_x)_\nu}{\omega}=
\frac{|\varepsilon_{\text{med},x}|^2 \, \im \Pi^{11}}{m_{A^\prime}}\,.
\end{equation}
Absorption rates for the other two polarizations are obtained in complete analogy. One simply has to replace $\Pi^{11}$ by $\Pi^{22}$ ($\Pi^{33}$) for $y$-polarized ($z$-polarized) dark photons.

In principle the incoming dark photon polarization needs to be evaluated in the laboratory frame. This complication is, however, usually irrelevant since the dark photons in the solar neighborhood are expected to be unpolarized. Therefore, the rate is simply given by the average
\begin{equation}
 \Gamma = \frac{1}{3} \left(\Gamma_x+\Gamma_y +\Gamma_z \right)=\frac{|\varepsilon_{\text{med},x}|^2 \, \im \Pi^{11}+|\varepsilon_{\text{med},y}|^2 \, \im \Pi^{22}+|\varepsilon_{\text{med},z}|^2 \, \im \Pi^{33}}{3\,m_{A^\prime}}\,.
 \label{eq:absorption}
\end{equation}
The total absorption rate in the detector per unit mass is obtained as
\begin{equation}
 R_{\text{tot}} = \frac{\rho_{A^\prime}}{\rho_T\,m_{A^\prime}} \,\Gamma\,.
\end{equation}
The lowest dark photon mass which can be probed by a Dirac material is set by the band gap. Furthermore, the rate has to be cut off when the largest energy deposit consistent with the linear dispersion relation is reached at $m_{A^\prime}=2\tilde{\Lambda}$~\cite{hochberg}.

We emphasize that the absorption of unpolarized dark photons is time-independent. This is because the spatial components of the polarization tensor $\Pi^{ii}$ (with $i=1,2,3$) are independent of the three-momentum transfer (in the relevant limit $\omega\gg |\mathbf{q}|$). This statement disagrees with Ref.~\cite{Coskuner:2019odd} which found a large daily modulation in anisotropic Dirac materials. The discrepancy arises because the photon polarization tensor employed in Ref.~\cite{Coskuner:2019odd} carries a residual $\tilde{\mathbf{q}}^2$-dependence which would favor scattering in the direction of the largest Fermi velocity. We will show in the next section that such a momentum dependence is absent and that the absorption rate remains constant with time.

\section{Polarization Tensor in Dirac Materials}\label{sec:polarizationtensor}

In this section, we will derive the photon polarization tensor for Dirac materials. The Lagrangian describing photons and electronic excitations in Dirac materials reads
\begin{equation}
 \mathcal{L} = -\frac{1}{4} F_{\mu\nu}F^{\mu\nu} + i\,\bar{\psi}\tilde{\gamma}^\mu(\partial_\mu+i e A_\mu)\psi - \Delta \bar{\psi}\psi\,.
\end{equation}
For convenience, we introduced the rescaled gamma matrices 
\begin{equation}\label{eq:gamma}
  \tilde{\gamma}^\mu = \{\gamma_0,v_{\mathrm{F},x} \gamma_1,v_{\mathrm{F},y} \gamma_2,v_{\mathrm{F},z} \gamma_3\}\,.
\end{equation}
Compared to the Lagrangian of quantum electrodynamics, the speed of light is replaced by the Fermi velocity in the corresponding spatial direction. Furthermore, the role of the electron mass term is played by $\Delta$ which is half the band gap. Notice that the structures of the electron kinetic term and the electron-photon vertex coincide as required by gauge invariance. 

\begin{figure}[htp]
\begin{center}   
 \includegraphics[width=4.5cm]{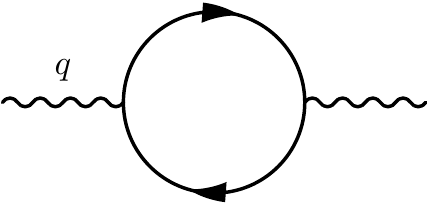}
\end{center}
\caption{Feynman diagram for the polarization tensor}
\label{fig:polarization}
\end{figure}

At first order in perturbation theory, the photon polarization tensor is obtained from the diagram shown in Fig.~\ref{fig:polarization}. The corresponding amplitude reads
\begin{equation}
 \Pi^{\mu}_{\phantom{\mu}\nu}(q) = -\frac{(-ie)^2}{\kappa}\frac{g}{2} \int \frac{\mathrm{d}^4 k}{(2\pi)^4} \text{Tr}\left\{\tilde{\gamma}^\mu\frac{i}{\tilde{\cancel k}-\Delta}\tilde{\gamma}_\nu \,\frac{i}{\tilde{\cancel k}+\tilde{\cancel q}-\Delta} \right\}\,,
\end{equation}
where the rescaled four-momenta $\tilde{q}^\mu$ and $\tilde{k}^\mu$ are defined analogous to $\tilde{\gamma}^\mu$ in eq.~\eqref{eq:gamma}. The case of a single Dirac fermion in the loop corresponds to a single Dirac cone ($g_C=1$) with spin-degeneracy $g_s=2$. We keep the factor $g$ in the above expression in order to allow for generic Dirac cone degeneracy. The background dielectric constant $\kappa$ is taken to be isotropic for the moment (the case of anisotropic $\kappa$ will be discussed below). In order to employ the well-known result for the vacuum polarization tensor in QED (see e.g.\ Ref.~\cite{Peskin:1995ev}), it is convenient to transform the integration measure from $k$ to $\tilde{k}$. Furthermore, we need to regularize the integral. Choosing the dimensional regularization scheme, we  perform the following replacement
\begin{equation}
 \mathrm{d}^4 k \rightarrow (v_{\mathrm{F},x}\,v_{\mathrm{F},y}\,v_{\mathrm{F},z})^{-1}\,\tilde{\mu}^{4-d} \mathrm{d}^d \tilde{k}\,,
\end{equation}
where $\tilde{\mu}$ denotes the renormalization scale. The resulting polarization tensor can be written in the form
\begin{equation}\label{eq:tensor1}
 \Pi^{\mu}_{\phantom{\mu}\nu}(q) = 
 \begin{pmatrix}
                                     -q_x^2 v_{\mathrm{F},x}^2-q_y^2 v_{\mathrm{F},y}^2-q_z^2 v_{\mathrm{F},z}^2 & \omega q_x v_{\mathrm{F},x}^2 & \omega q_y v_{\mathrm{F},y}^2 & \omega q_z v_{\mathrm{F},z}^2 \\[1mm] -\omega q_x v_{\mathrm{F},x}^2 & \omega^2 v_{\mathrm{F},x}^2 & 0 & 0 \\[1mm] -\omega q_y v_{\mathrm{F},y}^2 & 0 & \omega^2 v_{\mathrm{F},y}^2 & 0 \\[1mm] -\omega q_z v_{\mathrm{F},z}^2 & 0 & 0 & \omega^2 v_{\mathrm{F},z}^2
                                    \end{pmatrix}
                                    \frac{\pi(\tilde{q}^2)}{\kappa}\,,
\end{equation} 
where we neglected quartic terms in the $v_{\mathrm{F},i}$. This is justified since the Fermi-velocities are much smaller than the speed of light. For later convenience, we have not included $\kappa$ in the definition of $\pi(\tilde{q}^2)$. As a consistency check, one can easily verify that the polarization tensor fulfills the Ward identities $\Pi^{\mu}_{\phantom{\mu}\nu} q_\mu=\Pi^{\mu}_{\phantom{\mu}\nu} q^\nu=0$. This implies that the photon remains massless within the Dirac material. Let us now turn to the polarization function. We find
\begin{equation}
 \pi(\tilde{q}^2) = -\frac{4 g\, e^2  }{(4\pi)^{d/2}\,v_{\mathrm{F},x} v_{\mathrm{F},y} v_{\mathrm{F},z}} \int\limits_0^1 \mathrm{d}x\; x\, (1-x)\, \Gamma\left(2-\tfrac{d}{2}\right)\left(\frac{\tilde{\mu}^2}{\Delta^2-x(1-x)\tilde{q}^2}\right)^{2-d/2}
\end{equation}
In the modified minimal subtraction scheme ($\overline{\text{MS}}$), one replaces
\begin{equation}
 \frac{\Gamma\left(2-\tfrac{d}{2}\right)}{(4\pi)^{d/2}}A^{2-d/2}\;\;\longrightarrow\;\; \frac{1}{(4\pi)^2} \log A\,,
\end{equation}
and hence
\begin{equation}
 \pi(\tilde{q}^2)= -\frac{g\,e^2}{4\pi^2 \,v_{\mathrm{F},x} v_{\mathrm{F},y} v_{\mathrm{F},z}} \int\limits_0^1 dx\; x\, (1-x)\, \log\left(\frac{\tilde{\mu}^2}{\Delta^2-x(1-x)\tilde{q}^2}\right)\,.
\end{equation}

In the following, we set the renormalization scale to the cutoff $\tilde{\Lambda}$. This choice is motivated by the matching condition for the effective electron charge which is given by $e/\sqrt{\kappa}$ at the cutoff (where electrons should behave as in an insulator). In order to recover the standard expression of $\Pi^{00}$ for an isotropic Dirac material with vanishing band gap (see e.g.~\cite{thakur}), the precise replacement is
\begin{equation}
\tilde{\mu}\rightarrow 2\, e^{-5/6}\,\tilde{\Lambda}
\end{equation}
The imaginary part of the polarization function, which arises from a negative argument in the logarithm, can be evaluated analytically. One finds
\begin{align}\label{eq:polarizationfunction}
 \pi(\tilde{q}^2)= -\frac{g\,e^2}{4\pi^2 \,v_{\mathrm{F},x} v_{\mathrm{F},y} v_{\mathrm{F},z}} &\left[
 \int\limits_0^1 dx\; x\, (1-x)\, \log\left|\frac{4 \,e^{-5/3}\,\tilde{\Lambda}^2}{\Delta^2-x(1-x)\tilde{q}^2}\right| \right.\nonumber\\
 &\left.\phantom{\int\limits_0^1}+ \frac{i\pi}{6} \left(1 + \frac{2\Delta^2}{\tilde{q}^2}\right)\; \sqrt{1 - \frac{4\Delta^2}{\tilde{q}^2}}\;\;\Theta\left(\tilde{q}^2-4\Delta^2\right)
  \right]\;.
\end{align}
For convenience we also state the result for vanishing band gap,
\begin{equation}\label{eq:polarizationfunction2}
 \pi(\tilde{q}^2)= -\frac{g\,e^2}{24\pi^2 \,v_{\mathrm{F},x} v_{\mathrm{F},y} v_{\mathrm{F},z}} \left(
 \log\left|\frac{4\tilde{\Lambda}^2}{\tilde{q}^2}\right|+i\pi\Theta\left(\tilde{q}^2\right)\right)\,.
\end{equation}
We finally want to generalize the photon polarization tensor to the case of an anisotropic background dielectric tensor. Along the principal axes, the latter can be chosen diagonal such that we have 
\begin{equation}
  \kappa = \begin{pmatrix}
    \kappa_{xx} & 0 & 0 \\
    0 & \kappa_{yy} & 0 \\
    0 & 0 & \kappa_{zz}
  \end{pmatrix}\;.
\end{equation}
Given this form, the spatial components of the polarization tensor can be obtained by the replacement $\Pi_{ii}/\kappa \rightarrow \Pi_{ii}/\kappa_{ii}$ in~eq.~\eqref{eq:tensor1}~\cite{hochberg}. The remaining components are fixed by the Ward identities. The most general expression for the polarization tensor thus reads
\begin{equation}\label{eq:tensor2}
  \Pi^{\mu}_{\phantom{\mu}\nu}(q) = 
 \begin{pmatrix}
                                     -q_x^2 \frac{v_{\mathrm{F},z}^2}{\kappa_{xx}}-q_y^2 \frac{v_{\mathrm{F},y}^2}{\kappa_{yy}}-q_z^2 \frac{v_{\mathrm{F},z}^2}{\kappa_{zz}} & \omega q_x \frac{v_{\mathrm{F},x}^2}{\kappa_{xx}} & \omega q_y \frac{v_{\mathrm{F},y}^2}{\kappa_{yy}} & \omega q_z \frac{v_{\mathrm{F},z}^2}{\kappa_{zz}} \\[1mm] -\omega q_x \frac{v_{\mathrm{F},x}^2}{\kappa_{xx}} & \omega^2 \frac{v_{\mathrm{F},x}^2}{\kappa_{xx}} & 0 & 0 \\[1mm] -\omega q_y \frac{v_{\mathrm{F},y}^2}{\kappa_{yy}} & 0 & \omega^2 \frac{v_{\mathrm{F},y}^2}{\kappa_{yy}} & 0 \\[1mm] -\omega q_z \frac{v_{\mathrm{F},z}^2}{\kappa_{zz}} & 0 & 0 & \omega^2 \frac{v_{\mathrm{F},z}^2}{\kappa_{zz}}
                                    \end{pmatrix} 
 \, \pi(\tilde{q}^2)\,.
\end{equation}
Notice that in the kinematic regime relevant for DM scattering $|\mathbf{q}|\gg\omega$, the polarization tensor is strongly dominated by the $\Pi^{00}$-component. With this simplification, the Schwinger-Dyson equation leads to the photon propagator of eq.~\eqref{eq:propagatorscatter} and therefore
\begin{equation}
 \mathcal{F}_{\text{med}}(q)= \frac{1}{1 + \left(q_x^2 \frac{v_{\mathrm{F},z}^2}{\kappa_{xx}} + q_y^2 \frac{v_{\mathrm{F},y}^2}{\kappa_{yy}} + q_z^2 \frac{v_{\mathrm{F},z}^2}{\kappa_{zz}}\right) \frac{\pi(\tilde{q}^2)}{q^2}} \; .
 \label{eq:Fmed}
\end{equation}
This expression improves the corresponding expression in Ref.~\cite{hochberg}, where the geometric mean of the components of $\kappa$ is taken instead of including them individually.

In the opposite regime $\omega\gg|\mathbf{q}|$ which is relevant for dark photon absorption, the spatial components of $\Pi$ dominate and one obtains the photon propagator of eq.~\eqref{eq:propagatorabsorb}. We make the important observation that for $\omega\gg|\mathbf{q}|$, $\Pi^{ij}$ becomes independent of the three-momentum $\mathbf{q}$ (since the function $\pi(\tilde{q}^2)$ in~\eqref{eq:tensor2} only depends on $\omega^2$ in this regime). As stated earlier, this implies that the dark photon absorption rate in Dirac materials does not depend on the direction of the momentum transfer.

\section{Candidate Dirac Materials}
\label{sec:candidates}
For our study we consider three potential candidates for Dirac materials based DM sensors: ZrTe$_5$, Yb$_3$PbO, and BNQ-TTF. In this section we present calculations of their respective band structures and determine the relevant properties. The ab initio calculations were performed in the framework of the density functional theory (DFT) using a pseudopotential projector augmented-wave method~\cite{hamann1979norm,blochl1994projector,pseudo1,pseudo2}, as implemented in the Vienna Ab initio Simulation Package (VASP)~\cite{vasp2,kresse1999ultrasoft}. We compare results for the experimental crystal structure (NR) with results from structurally optimized crystal structures, which where obtained by allowing the unit cell volume to change, but keeping the unit cell shape and the atomic positions unchanged (ISIF7). For the structural optimization and the band structure calculations, we have used the semilocal meta-GGA functional (SCAN) \cite{sun2015,sun2016accurate}. To get reliable optimized structural ground states we added Van der Waals corrections according to Tkatchenko and Scheffler \cite{tkatchenko2009accurate} for the calculations concerning ZrTe$_5$ and BNQ-TTF. 

\begin{figure}[b!]
\scriptsize
    \centering
    \subfloat[]{\includegraphics[height=4.5cm]{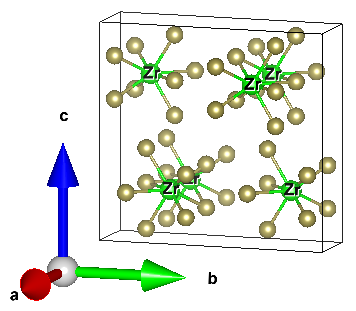}}\hspace{0.5cm}
    \subfloat[]{\includegraphics[height=4.5cm]{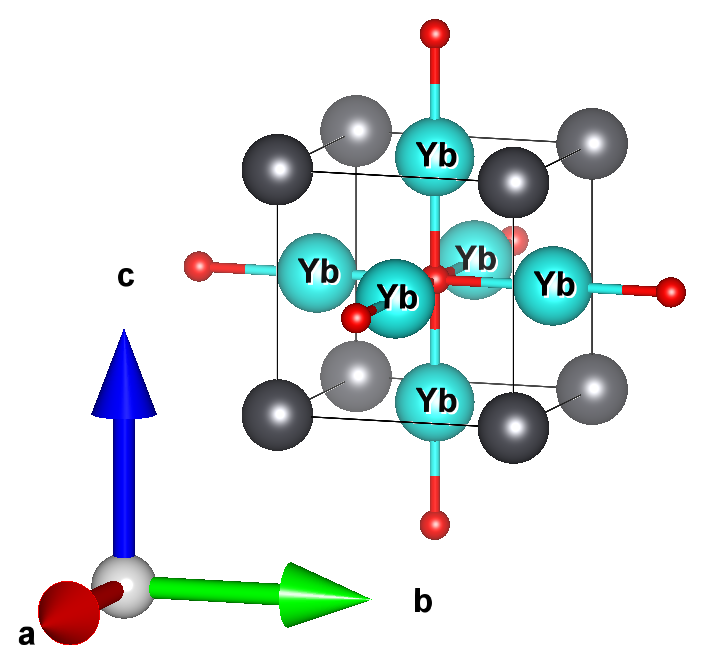}}\hspace{0.5cm}
    \subfloat[]{\includegraphics[height=4.5cm]{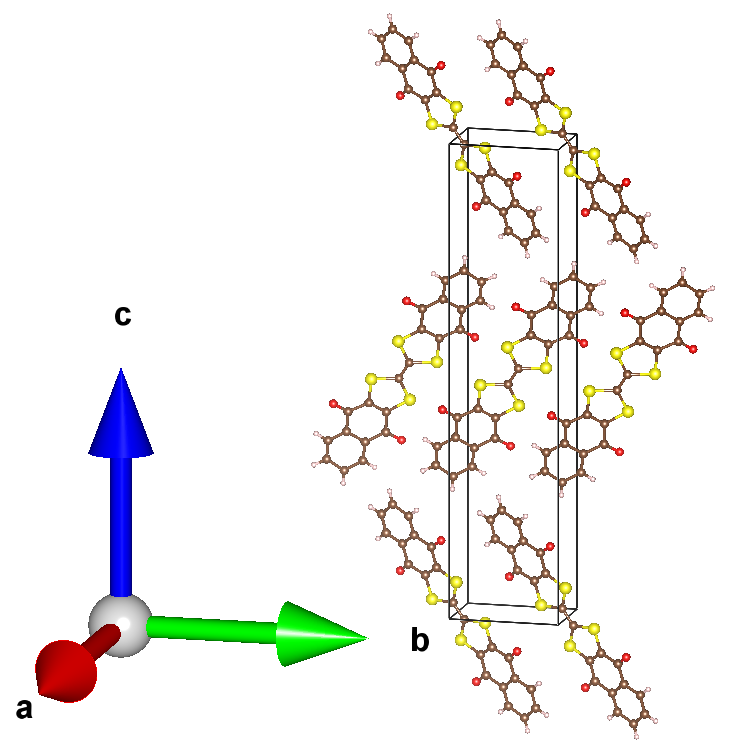}}\\
    \subfloat[]{
    \begin{tabular}{lllccccccccc}
    \hline\hline 
    Material & Space Group &  & a & b & c & $\alpha$ & $\beta$ & $\gamma$ & $V_{\text{UC}}$  & $\rho$  & $n_e$ \\
     &  &  & (\AA)  &  (\AA)  &  (\AA)  &  &  &  & (\AA$^3$) & (gcm$^{-3}$) & ($10^{23}$kg$^{-1}$)\\
    \hline
      ZrTe$_5$& Cmcm (63) & Exp. \cite{zrte5} & 3.987 & 14.53 & 13.722 & 90 & 90 & 90 & 795.146 & 6.089 & 2.065\\
              & & this work &  3.978 & 14.494 & 13.690 & 90 & 90 & 90 & 789.292 & 6.135 & 2.065\\
      Yb$_3$PbO & Pm$\overline{3}$m (221) & Exp. \cite{velden2004kenntnis} & 4.859 & 4.859 & 4.859 & 90 & 90 & 90 & 114.700 & 10.744 & 8.115\\
              & & this work &  4.737 & 4.737 & 4.737 & 90 & 90 & 90 & 106.272 & 11.596 & 8.115 \\
     BNQ-TTF & P2$_1$/n (14) & Exp. \cite{BNQ}& 3.881 & 7.532 & 31.350 & 90 & 96.476 & 90 & 916.365 & 1.683 & 6.480 \\
      & & this work & 3.899 & 7.567 & 31.698 & 90 & 96.476 & 90 & 929.339 & 1.659 & 6.480 \\
           \hline\hline 
    \end{tabular}}
 \caption{Crystal structure information of the considered Dirac materials. (a)-(c) show the unit cells ZrTe$_5$, Yb$_3$PbO, and BNQ-TTF. (d) Experimental and computational lattice constants, unit cell volumes, and densities. The electron density $n_e$ specifies the density obtained for a single electron per unit cell.\label{structure}}
        \centering
\end{figure}

For the $\vec{k}$-space integration, we chose a $\Gamma$-centered mesh~\cite{monkhorst1976special} with 14$\times$4$\times$4 points for ZrTe$_5$, $10\times10\times10$ points for Yb$_3$PbO, and 14$\times$8$\times$2 points for BNQ-TTF. The cut-off energy was set to 600 eV. The calculation of the dielectric tensor was performed using the generalized gradient approximation according to Perdew, Burke, and Ernzerhof \cite{perdew1996} and density functional perturbation theory. The calculations for the band structure and dielectric tensor were performed with spin-orbit coupling, the structural optimization was done without spin-orbit coupling. For Yb$_3$PbO the $f$-electrons are considered to be occupied. To push related electronic bands occurring at the Fermi level into the valence band we applied the GGA+Hubbard-U correction using a value of $U=10$ eV for the Yb-$f$-orbitals as suggested in Ref. \cite{pertsova2019}.

The unit cells and obtained lattice parameters from the structural optimization in comparison with the reported experimental lattice constants are shown in Fig.~\ref{structure}. We observe that the overall unit cell volume for the computational ground state is slightly decreased for ZrTe$_5$ and Yb$_3$PbO and slightly increased for BNQ-TTF. The increase of the unit cell volume for the structural ground state for organic materials is common and can be traced back to a slightly increased bond length occurring in the DFT calculations.

\begin{figure*}[t!]
    \centering
    \includegraphics[width=0.32\textwidth]{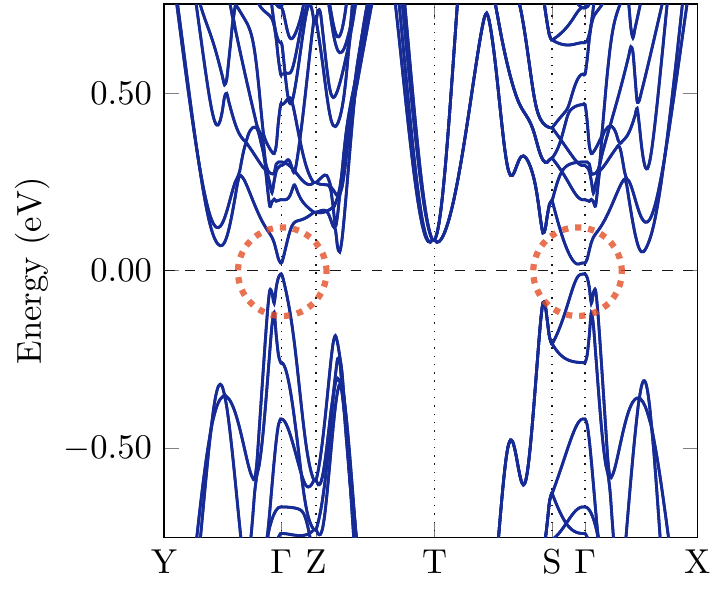}
    \includegraphics[width=0.32\textwidth]{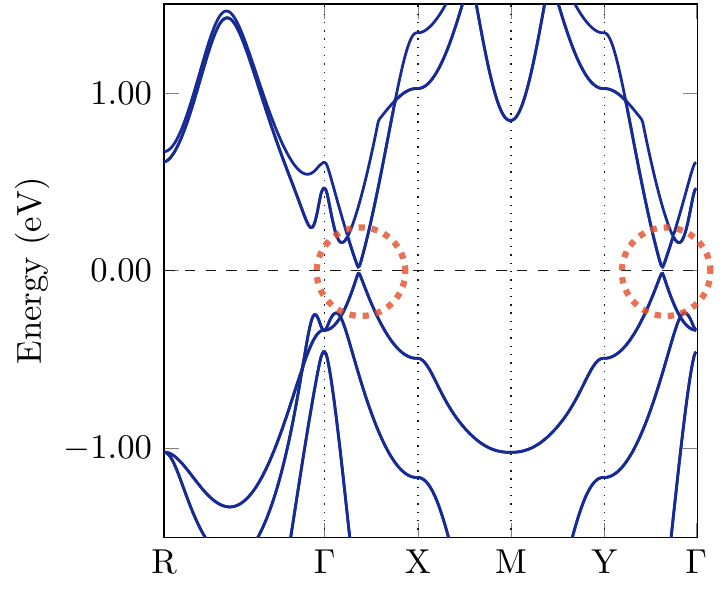}
    \includegraphics[width=0.32\textwidth]{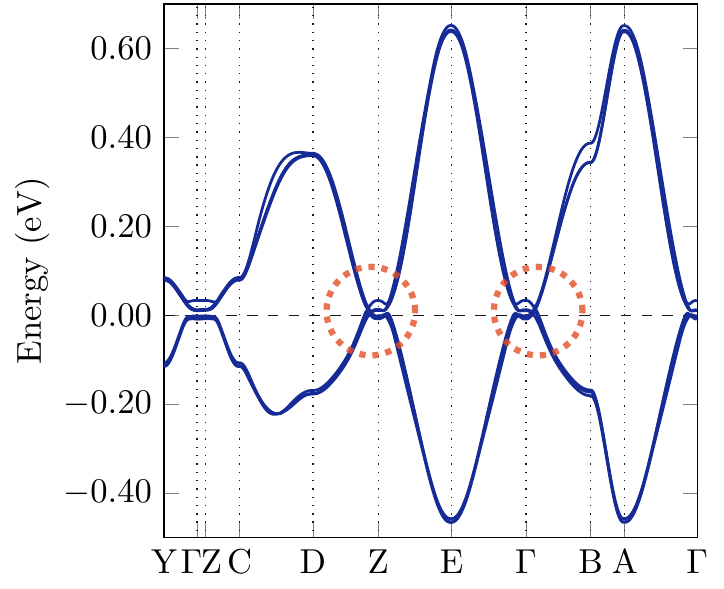}
    \caption{Calculated ab initio band structures for potential Dirac DM sensor materials: ZrTe$_5$ (left), Yb$_3$PbO (center), and BNQ-TTF (right).}
    \label{bandsDM4DM}
\end{figure*}

The obtained band structures for ZrTe$_5$, Yb$_3$PbO, and BNQ-TTF are shown in Fig.~\ref{bandsDM4DM}. ZrTe$_5$ exhibits a gaped Dirac point at $\Gamma$, the center of the Brillouin zone. The calculated band gap with and without structural optimization are given by 31.2 meV and 23.6 meV, which corresponds to $\Delta$ = 15.6 meV and $\Delta$ = 11.8 meV, respectively. Yb$_3$PbO exhibits a gaped Dirac point along the path $\overline{\Gamma X}$ ($\Gamma=(0,0,0)$, $X=(0.5,0,0)$) located at $\vec{k}_D = (0.185,0.0,0.0)$. The corresponding band gap is 34.4 meV ($\Delta$ = 17.2 meV) for the experimental unit cell and 38.8 meV ($\Delta$ = 19.4 meV) for the optimized unit cell. Due to the cubic symmetry of the system a total of 6 such points can be observed which can be projected by applying 4-fold rotations about the $k_y$- and $k_z$- axis in the Brillouin zone. 

\begin{table}[t!]
\scriptsize
    \centering
    \begin{tabular}{cccccccccc}
    \hline\hline
        Material & Mode & $v_{\mathrm{F},x}$ & $v_{\mathrm{F},y}$ & $v_{\mathrm{F},z}$ & $\Delta$ & $\Lambda_x$ & $\Lambda_y$  & $\Lambda_z$  & $\vec{k}_{\text{cone}}$  \\
         &  & (c) & (c) & (c) & (meV) & (\AA$^{-1}$) & (\AA$^{-1}$) &  (\AA$^{-1}$) &  \\
        \hline
        ZrTe$_5$ & NR & 1.1$\times 10^{-3}$  & 4.5$\times 10^{-4}$ & 1.0$\times 10^{-3}$ & 11.8 & 0.23  & 0.215 & 0.1 & (0.,0.,0.)\\
        & \bf ISIF7 & $\mathbf{1.1\times 10^{-3}}$  & $\mathbf{4.4\times 10^{-4}}$ & $\mathbf{9.1\times 10^{-4}}$ & \bf 15.6 & \bf 0.23 & \bf 0.216 & \bf 0.1 & \bf (0.,0.,0.)\\
       & Th.  \cite{hochberg} & 2.9$\times 10^{-3}$  & 5.0$\times 10^{-4}$ & 2.1$\times 10^{-3}$ & 17.5 & 0.07 & 0.07 & 0.07 & (0.,0.,0.) \\
       & Exp. \cite{zheng2016} & 1.3$\times 10^{-3}$  & 6.5$\times 10^{-4}$ & 1.6$\times 10^{-3}$ & 11.75 & & & & (0.,0.,0.) \\
        \hline
        Yb$_3$PbO & NR & 8.5$\times 10^{-4}$& 8.8$\times 10^{-4}$ & 8.8$\times 10^{-4}$ & 17.2  & 0.45 & 0.45 & 0.45 & ($\pm$0.185,0.,0.0)\\
         &  & 8.8$\times 10^{-4}$ & 8.5$\times 10^{-4}$ & 8.8$\times 10^{-4}$ &   & & & & (0.0,$\pm$0.185,0.0) \\
         &  &  8.8$\times 10^{-4}$ & 8.8$\times 10^{-4}$ & 8.5$\times 10^{-4}$&   & & &  & (0.0,0.0,$\pm$0.185) \\
        & ISIF7 & 8.7$\times 10^{-4}$& $9.0\times 10^{-4}$ & $9.0\times 10^{-4}$ & 19.4  & 0.45 & 0.45 & 0.45 & ($\pm$0.185,0.,0.0) \\
        &  &  $9.0 \times 10^{-4}$ & 8.7$\times 10^{-4}$& $9.0 \times 10^{-4}$ & & & &  & (0.0,$\pm$0.185,0.0) \\
        &  & $9.0\times 10^{-4}$ & $9.0\times 10^{-4}$ & 8.7$\times 10^{-4}$&  & & & & (0.0,0.0,$\pm$0.185) \\
        & \bf Impl.  & $\mathbf{8.9\times 10^{-4}}$ & $\mathbf{8.9\times 10^{-4}}$ & $\mathbf{8.9\times 10^{-4}}$ & \bf 19.4  & \bf 0.45 & \bf 0.45 & \bf 0.45 & \\
        \hline
        BNQ-TTF & NR & 2.3 $\times 10^{-4}$ & 1.9 $\times 10^{-4}$ & - & 0. & 0.81 & 0.3 & 0.1 &  ($\pm$0.075,0.,0.5) \\
         &  & 1.9 $\times 10^{-4}$ & 2.3 $\times 10^{-4}$ & - & & 0.3 & 0.81 & 0.1 &  ($\pm$0.075,0.,0.) \\
        & ISIF7 & 2.2 $\times 10^{-4}$ & 1.8 $\times 10^{-4}$ & - & 0. & 0.81 & 0.3 & 0.1 &  ($\pm$0.065,0.,0.5)  \\
        &  & 1.8 $\times 10^{-4}$ & 2.2 $\times 10^{-4}$ & - & & 0.3 & 0.81 & 0.1 &  ($\pm$0.065,0.,0.)  \\
        & \bf Impl. & $\mathbf{2.0 \times 10^{-4}}$ & $\mathbf{2.0 \times 10^{-4}}$ & $\mathbf{1.0 \times 10^{-4}}$  & \bf 5.0 & \bf 0.3 & \bf 0.3 & \bf 0.1 &   \\
     \hline\hline
    \end{tabular}
    \caption{Calculated Fermi-velocities, band gaps, cut-off radii, and Dirac point positions in the Brillouin zone. We compare values obtained using optimized structures (ISIF7) and experimental structures (NR). The values implemented for our sensitivity estimates are highlighted in bold.}
    \label{diracvelocities}
\end{table}

In Ref. \cite{geilhufe2018materials}, BNQ-TTF was discussed as a tiny gap organic semiconductor. However, our calculations incorporating spin-orbit coupling reveal a Dirac crossing along the paths $\overline{DZ}$ ($D=(0.5,0.0,0.5)$, $Z=(0.0,0.0,0.5)$) and $\overline{\Gamma B}$ ($B=(0.5,0.0,0.0)$) at $k_x = 0.075$ for the experimental and $k_x=0.065$ for the optimized unit cell. By two-fold rotational symmetry both points come with a partner with the corresponding values at $-k_x$. Organic materials are soft and therefore this material can be tuned by applying stress. A slightly strained sample of BNQ-TTF breaking the space group symmetries of the material is therefore likely to introduce a tiny gap. For our sensitivity estimates in the following section, we will consider a band gap of 10 meV ($\Delta = 5\,\mathrm{meV}$).

We furthermore performed additional band structure calculations for all three materials to fit the occurring Fermi velocities and determine the cut-off radii $\Lambda_i$ where the Dirac dispersion approximately holds. All values are summarized in Tab. \ref{diracvelocities}. The highest Fermi velocities are found for ZrTe$_5$ which are of the order of $10^{-3}$. In contrast, the flat bands of BNQ-TTF lead to very small Fermi velocities in the order of $10^{-4}$. Due to the low symmetry of ZrTe$_5$ and BNQ-TTF all three Fermi velocities come with different values. Furthermore, BNQ-TTF is a quasi 2-dimensional material were the dispersion in the $k_z$ direction of the Brillouin zone is extremely flat and the corresponding Fermi velocity vanishes. This effect is related to the weak hopping of electrons in the $c$-direction of the crystal stemming from the layered structure of the material. Applying pressure on the sample along the crystallographic $c$-direction will decrease the distance of molecules in the $c$-direction and therefore increase the hoping amplitudes between the molecules. As a result, an increased hopping amplitude will lead to a stronger dispersion of bands opening the opportunity to lift the flatness of the band and tune the Fermi velocity. In the following, we assume that in a sufficiently strained sample the flat direction will take a value of $v_{\mathrm{F},z} = 10^{-4}$ for our sensitivity estimates.\footnote{We note that for small Fermi velocities the effective coupling strength $\alpha_\text{eff} = \alpha / (\kappa \, v_{\mathrm{F}})$ increases and the material becomes increasingly strongly coupled. The considered Fermi velocities of BNQ-TTF imply $\alpha_\text{eff}\sim 10$. It is conceivable that perturbation theory still applies to systems with $\alpha_\text{eff}$ in this range (see discussion in~\cite{throckmorton}). Indeed, this has experimentally been verified for the case of graphene~\cite{Kotov}. Nevertheless, we wish to point out that our one-loop calculation of the polarization tensor should only be seen as qualitative estimate for the case of BNQ-TTF.
}

In comparison to ZrTe$_5$ and BNQ-TTF, Yb$_3$PbO crystallizes in a high-symmetry space group Pm$\overline{3}m$. As the Dirac point is observed, e.g., along the path $\overline{\Gamma X}$ the little group of $\vec{k}$ is given by $C_{4v}$ \cite{gtpack1,gtpack2}. Hence, the rotational symmetry enforces the two Fermi velocities corresponding to the directions orthogonal to $\overline{\Gamma X}$ to be degenerated, i.e., $v_{\mathrm{F},y}=  v_{\mathrm{F},z} \neq v_{\mathrm{F},x}$. For Yb$_3$PbO we observe slightly different values for $v_{\mathrm{F},i}$ in the conduction and the valence bands. In the conduction (valence) band, the value for $v_{\mathrm{F},x}$ is about $v_{\mathrm{F},x} \approx 2 v_{\mathrm{F},y}$ ($v_{\mathrm{F},x} \approx \frac{1}{2} v_{\mathrm{F},y}$). Hence the averaged values for $v_{\mathrm{F},x}$ given in Tab. \ref{diracvelocities} do not reflect this anisotropy. In the following we will use these averaged values to estimate the sensitivity of Yb$_3$PbO, but we will not attempt to calculate the modulation signal, which would require an extended formalism allowing for different Fermi velocities in the valence and conduction band.

\begin{table}[t]
\scriptsize
    \centering
    \begin{tabular}{ccccc|ccc|ccc|c}
    \hline\hline
        Material & Mode & $\kappa_{xx}$ & $\kappa_{yy}$ & $\kappa_{zz}$ & $\kappa_{xy}$ & $\kappa_{xz}$ & $\kappa_{yz}$  & $\kappa_{yx}$ & $\kappa_{zx}$ & $\kappa_{zy}$  & $g$ \\
        \hline
        ZrTe$_5$ & This work & 308.4 & 20.75 & 126.1 & -0.97 & -1.2 & -0.38 & 0.5 & -1.2 & -0.02 & 2 \\
        ZrTe$_5$ & Ref. \cite{hochberg} & 187.5  & 9.8 & 90.9 & & & & & &  \\
        \hline
        Yb$_3$PbO & This work & 42.8  & 42.8 & 42.8 & -12.38 & 8.58 & -12.38 & 8.58 & -12.38 & 8.58 & 12\\
        \hline
        BNQ-TTF & This work & 18.7 & 5.6 & 10.3  & -0.05 & 0.07 & -1. & -0.05 & 0.07 & -1. & 8\\
     \hline\hline
    \end{tabular}
    \caption{Dielectric tensor for ZrTe$_5$, Yb$_3$PbO, and BNQ-TTF calculated using density functional perturbation theory. In the final column we also specify the respective degeneracy $g = g_s \, g_C$.}
    \label{dielectric}
\end{table}

We furthermore calculated the values for the dielectric tensor by using density functional perturbation theory as implemented in the code VASP. The values are given in Tab. \ref{dielectric}. Due to the tiny gaps present in these materials these calculations are very sensitive to the gap size. However, we observe that for all three materials the diagonal elements $\kappa_{xx}$, $\kappa_{yy}$, and $\kappa_{zz}$ dominate over the off-diagonal components. The largest values are found for ZrTe$_5$ which is highly anisotropic with $\kappa_{xx}\approx 308$, but $\kappa_{yy}\approx 21$. The smallest values are seen for BNQ-TTF with $\kappa_{xx}\approx 19$ and $\kappa_{yy}\approx 6$. The cubic symmetry in Yb$_3$PbO enforces $\kappa_{xx}=\kappa_{yy}=\kappa_{zz}\approx 43$.

Finally, we need to determine the optimum orientation of the three Dirac materials in the laboratory. The coordinate system that we introduced above implies that the DM wind points in the $z$-direction at $t = 0\,\mathrm{days}$ and approximately in the $y$-direction at $t = 0.5\,\mathrm{days}$. We hence want to align the materials in such a way that the largest anisotropy is observed in the $y$-$z$ plane. In the following, we will always align the materials such that the smallest Fermi velocity points in the $y$ direction, while the largest Fermi velocity points in the $z$ direction.\footnote{For BNQ-TTF the two larger Fermi velocities are nearly degenerate. We align the detector such that the dielectric constant is smallest in the $z$ direction.} Since the event rate is largest when the DM wind is aligned with the smallest Fermi velocity, we expect a daily modulation that peaks at $t = 0.5 \, \mathrm{days}$.

\section{Sensitivity Estimates}
\label{sec:results}

We are now in the position to calculate the predicted DM signal as a function of time in the Dirac materials that we consider and to estimate their sensitivity. Before presenting our results, we first introduce the statistical approach that we employ.

\subsection{Statistical Method for Daily Modulation}

We will consider two possible outcomes for the experiments under consideration. First, we consider the case that the DM hypothesis is incorrect and that the experiments do not observe any DM signal. For example, if no events are observed at all, any parameter point predicting 3 or more events can be excluded at 95\% confidence level. If the experiment observes a number $N_\text{b}$ of background events, it can still exclude all parameter points for which the probability to observe at most $N_\text{b}$ signal events is less than 5\%.\footnote{A stronger bound can be obtained if a background model exists that would allow for background subtraction. Here we focus on the most conservative case in which no background model is assumed.}

The second outcome we consider is that the experiments do observe a DM signal. In this case it will be essential to confirm the DM nature of the excess by performing a test for daily modulation. Whether or not the daily modulation will be observable depends on both the amplitude of the modulation and the total (i.e.\ unmodulated) rate. We use the following approach to quantify the significance of the daily modulation.

Each day of observation is divided into the twelve hours around the expected maximum of the modulation and the twelve hours around the minimum of the modulation. Let $N_\text{max}$ be the total number of events that fall into the former window and $N_\text{min}$ the remaining events. Assuming $N_\text{max}, N_\text{min} \gg 1$ the event numbers are expected to follow a normal distribution with estimated standard deviation $\sqrt{N_\text{max}}$ and $\sqrt{N_\text{min}}$ respectively. Hence, the difference $N_\text{max} - N_\text{min}$ should follow a normal distribution with standard deviation $\sqrt{N_\text{max} + N_\text{min}}$. In the absence of a daily modulation, the expectation value of this quantity vanishes. To test the hypothesis that there is \emph{no} modulation, we can hence define the test statistic
\begin{equation}
 \chi_s^2 = \frac{(N_\text{max} - N_\text{min})^2}{N_\text{max} + N_\text{min}} \; ,
 \label{eq:chi2}
\end{equation}
which we have confirmed to follow a $\chi^2$ distribution with one degree of freedom under the null hypothesis using explicit Monte Carlo simulation. If $\chi_s^2 \gg 1$ there is positive evidence for a daily modulation and the hypothesis of no modulation can be rejected. For example, to reject the null hypothesis at 95\% confidence level, one would require $\chi_s^2 > 3.84$. More generally, the significance of the modulation is given by $\sqrt{\chi_s^2}$ standard deviations. 

In the following it will be useful to define the total number of signal events $N_\text{s} = N_\text{max} + N_\text{min}$ and the modulation fraction $A = (N_\text{max} - N_\text{min}) / (N_\text{max} + N_\text{min})$. With this definition, the test statistic can simply be written as $\chi_s^2 = A^2 N_\text{s}$. Hence, for a modulation fraction of $A = 20\%$ it is necessary to observe $N_\text{s} \approx 225$ events to detect $3\sigma$ evidence for a modulation, while for $A = 50\%$ fewer than 40 events may be sufficient. Note that given actual data, more sophisticated methods, such as a Lomb-Scargle~\cite{Lomb:1976wy,Scargle:1982bw} analysis, may reveal even higher significance for a modulation (see e.g.~\cite{Fox:2011px}).

Our approach is easily extended to include a number $N_\text{b}$ of background events. Assuming that the background does not modulate, it will cancel in the numerator but contribute to the denominator of eq.~(\ref{eq:chi2}), giving
\begin{equation}
 \chi_{sb}^2 = \frac{(N_\text{max} - N_\text{min})^2}{N_\text{max} + N_\text{min} + N_\text{b}} = \chi^2_s \frac{N_\text{s}}{N_\text{s}+N_\text{b}} \; . 
\end{equation}
We emphasize that this expression corresponds to the most conservative case without background subtraction and does not require any model of the expected background.

Let us consider the example of an unknown background which has a rate of 1 event per day. The total exposure is assumed to be 1 kg year. Based on the total number of observed events alone one can exclude all parameter points that would predict more than $\sim$400 signal events. Nevertheless, provided the modulation amplitude is sufficiently large, a substantially smaller number of signal events may be sufficient to identify a daily modulation. Indeed, given a modulation fraction of $50\%$ ($30\%$) it would only require about 135 (250) signal events to obtain $3\sigma$ evidence for daily modulation.

\subsection{Results for Dark Matter Scattering}
\begin{figure}[t!]
\begin{center}   
 \includegraphics[width=0.4\textwidth]{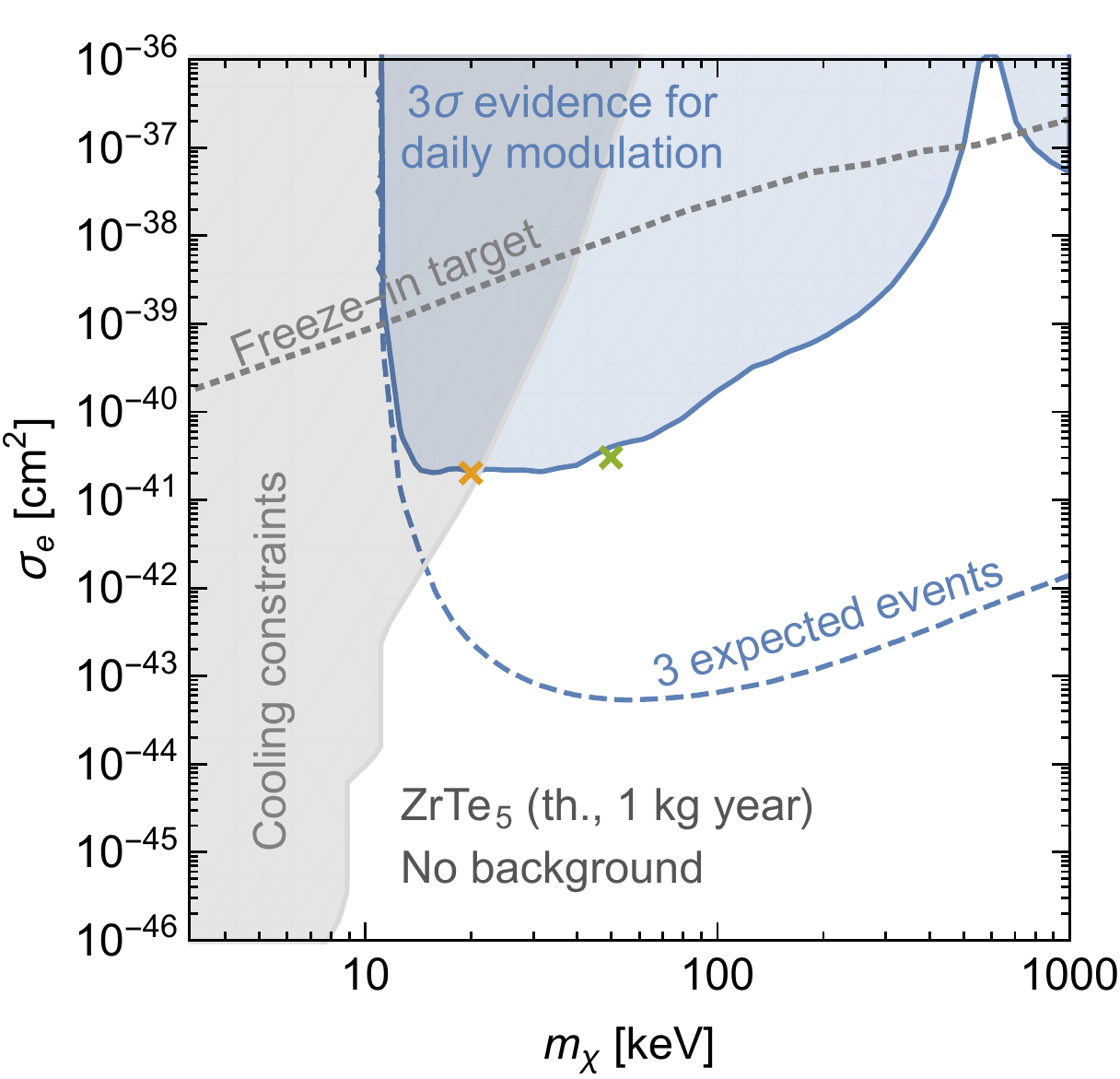}
  \includegraphics[width=0.4\textwidth]{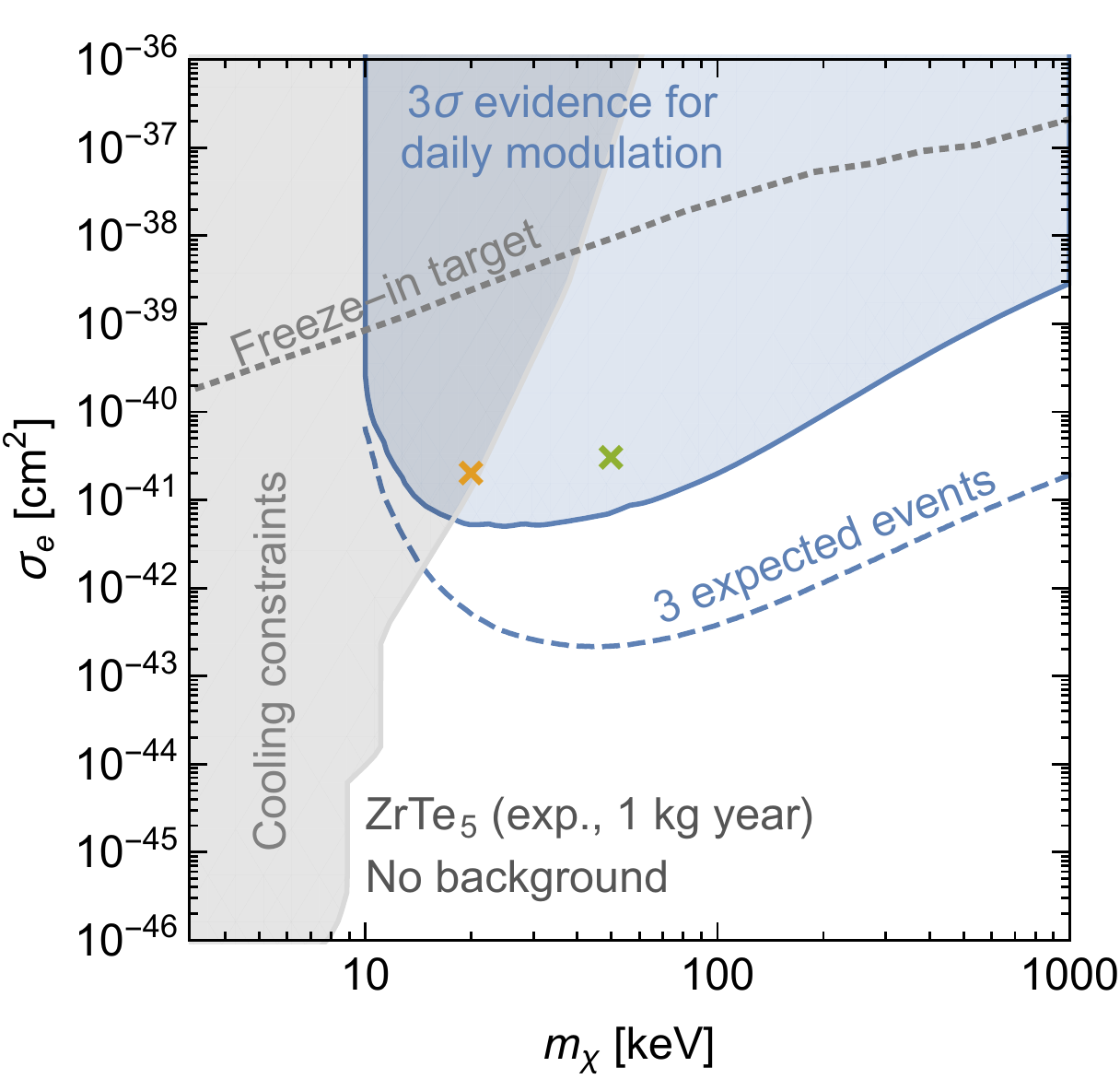}
  
  \includegraphics[width=0.4\textwidth]{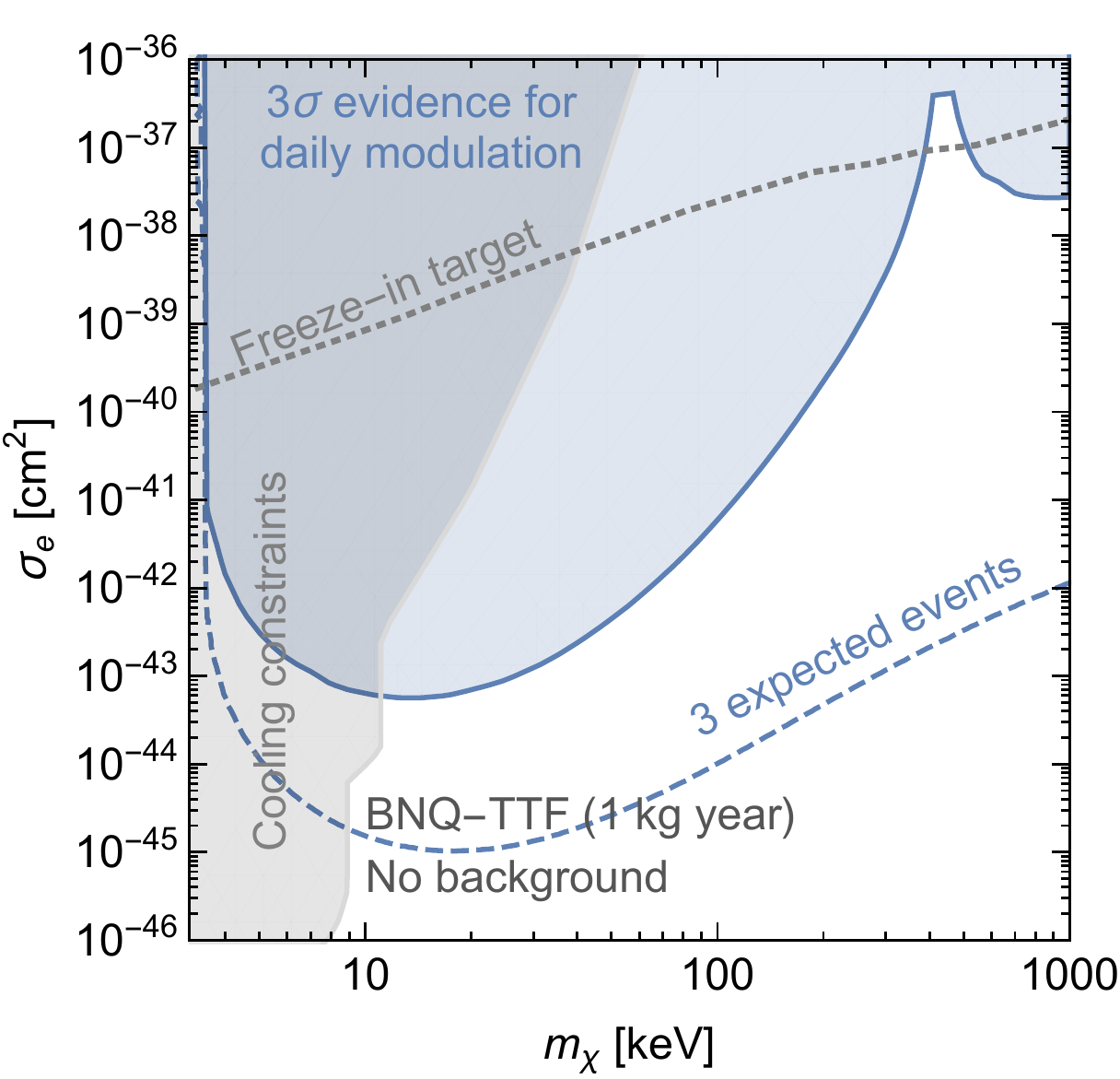}
  \includegraphics[width=0.4\textwidth]{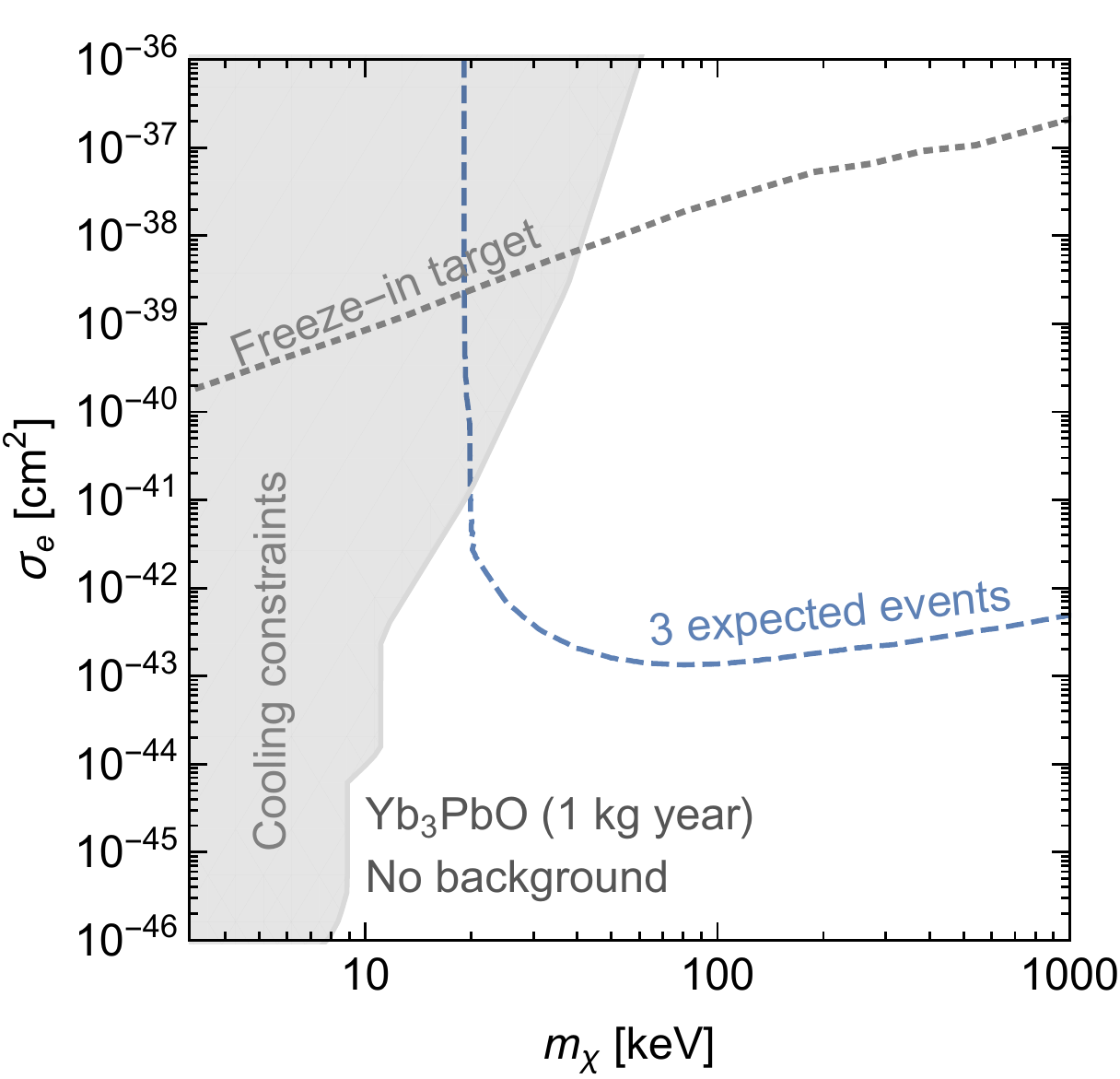}
\end{center}
\caption{Expected sensitivity for the Dirac materials ZrTe$_5$, BNQ-TTF and Yb$_3$PbO. The two panels in the top row correspond to different assumed properties for ZrTe$_5$ (see text for details). In the case of a null result, all parameter points above the dashed lines (corresponding to 3 expected events) can be excluded. In the shaded parameter region it will be possible to identify a daily modulation with $3\sigma$ significance in the case that a DM signal is observed.}
\label{fig:results}
\end{figure}

We present our main results in Fig.~\ref{fig:results} for three different Dirac materials. The two panels in the top row show the expected sensitivity for ZrTe$_5$, assuming the Fermi velocities and the band gap obtained from our calculations (left) and from experimental measurements (right). The two panels in the bottom row correspond to BNQ-TTF and Yb$_3$PbO, respectively. In each panel the dashed line indicates the exclusion bound from a null result, the shaded region in the first three panels indicates the parameter space where a daily modulation can be identified with $3\sigma$ significance. For the moment we assume that experimental backgrounds are negligible.

\begin{figure}[t]
\begin{center}   
\includegraphics[width=0.4\textwidth]{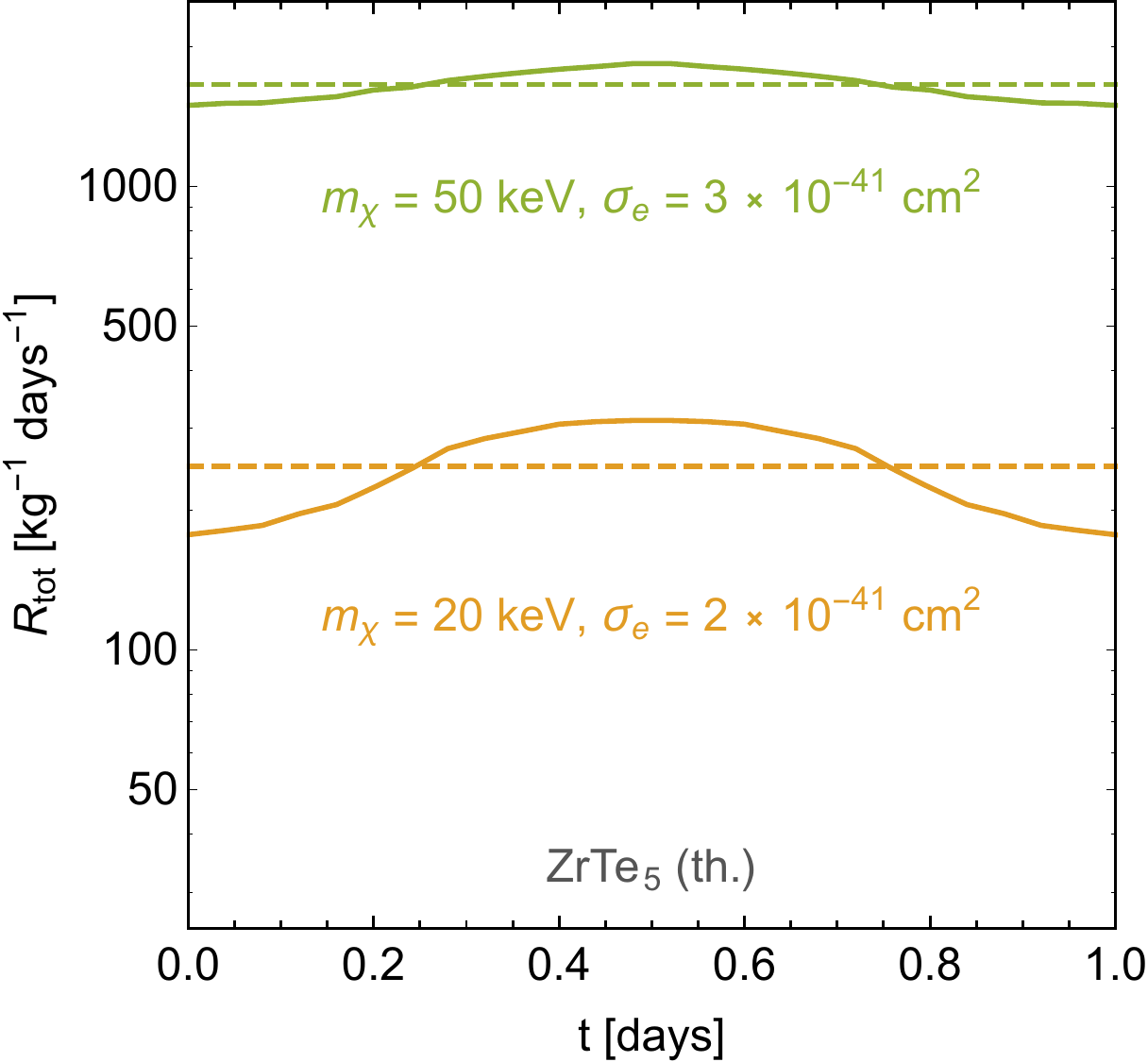}
\includegraphics[width=0.4\textwidth]{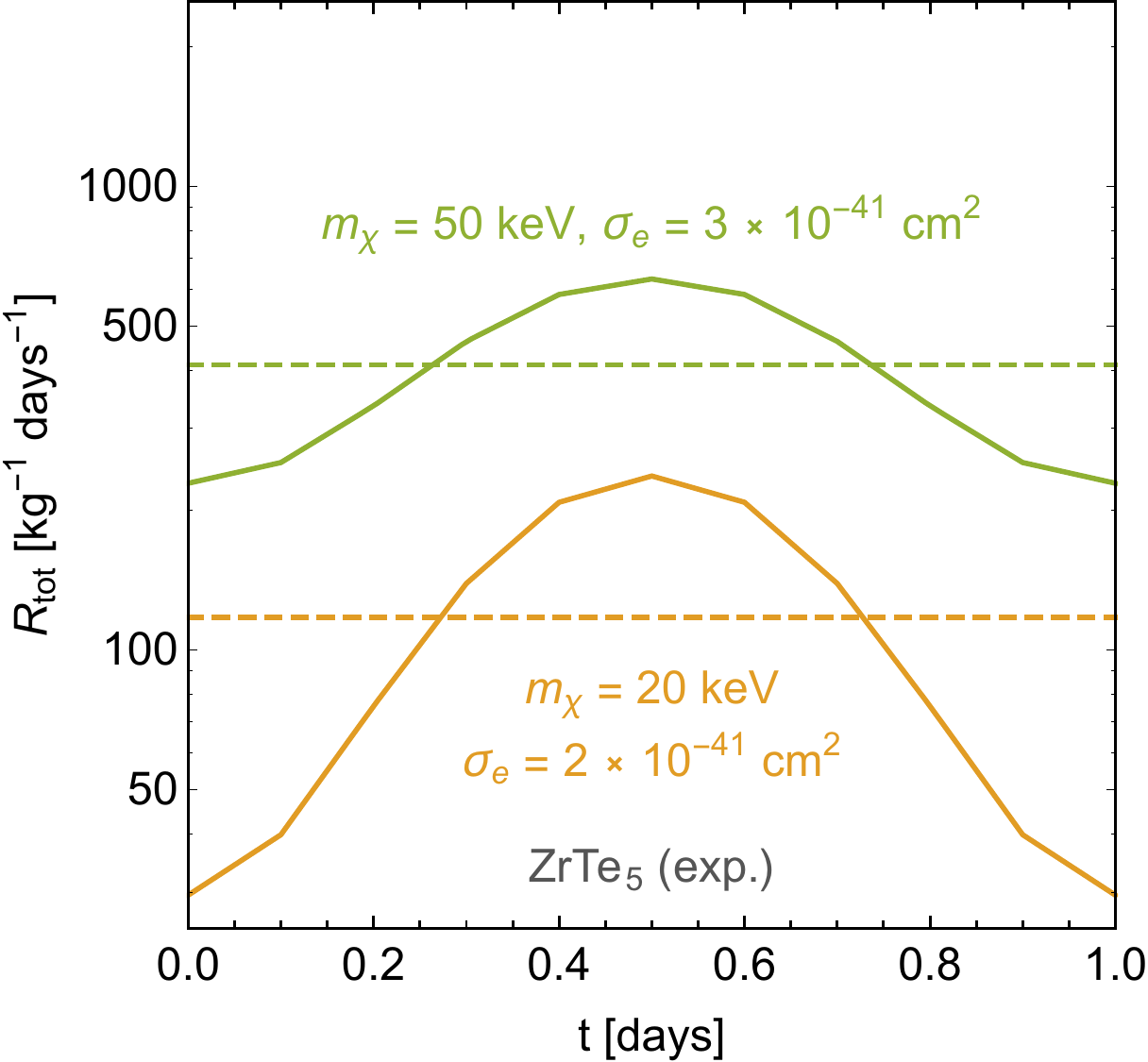}
  \end{center}
\caption{Event rate in ZrTe$_5$ as a function of time for the two benchmark points indicated in Fig.~\ref{fig:results}. The two panels correspond to different assumptions on the material properties.}
\label{fig:modulation}
\end{figure}

For comparison we show in each panel the combination of parameters for which the observed DM relic abundance can be reproduced via the freeze-in mechanism in a model with a massless dark photon. We include the contribution from plasmon decays, recently studied in Refs.~\cite{An:2018nvz,Dvorkin:2019zdi,Heeba:2019jho}. In the top row we furthermore indicate two benchmark points, corresponding to $m_\chi = 20\,\mathrm{keV}$, $\sigma_e = 2 \cdot 10^{-41}$ (orange) and $m_\chi = 50\,\mathrm{keV}$, $\sigma_e = 3 \cdot 10^{-41}$. The predicted event rate for these two points as a function of time is shown in Fig.~\ref{fig:modulation}.

\begin{figure}[t]
\begin{center}   
 \includegraphics[height=5cm]{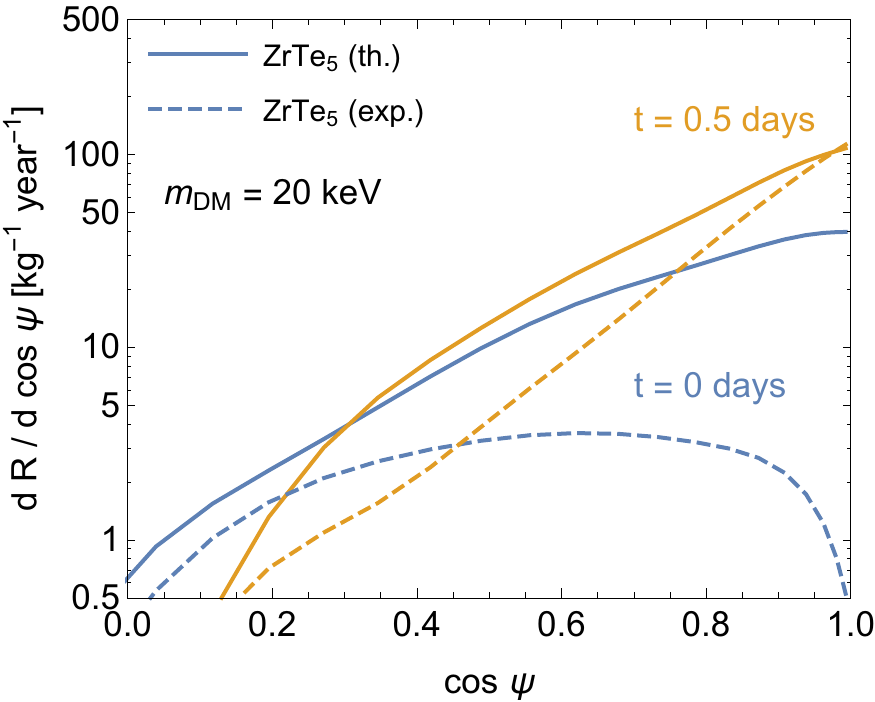}\qquad
  \includegraphics[height=5.2cm]{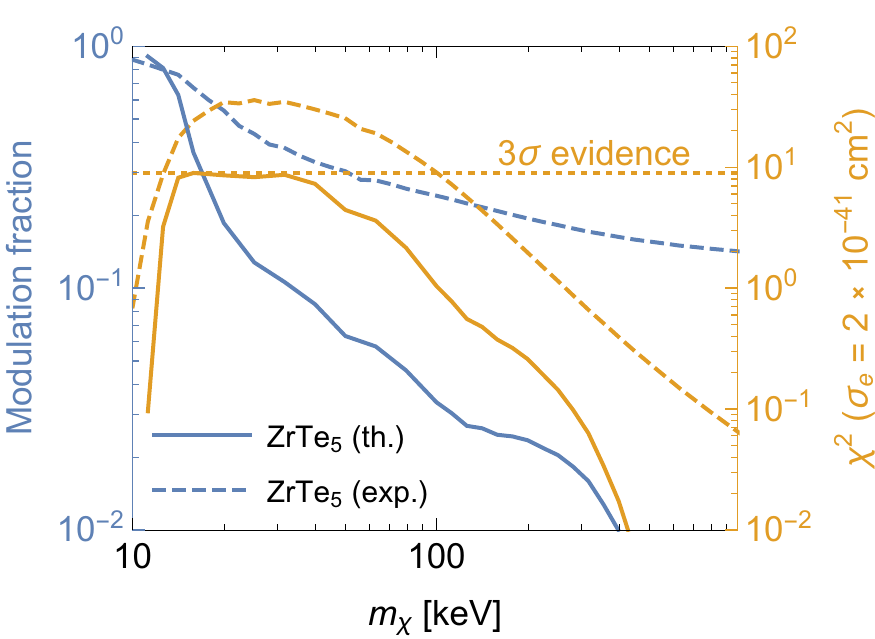}
\end{center}
\caption{Left: Differential event rate with respect to the cosine of the angle $\psi$ between the velocity of the Earth and the momentum transfer at $t = 0 \, \mathrm{days}$ (blue) and $t = 0.5 \, \text{days}$ (orange) for the theoretically calculated properties of ZrTe$_5$ (solid) and the experimentally measured properties (dashed). Right: Modulation amplitude (blue, left $y$-axis) and significance (orange, right $y$-axis) of a daily modulation for the two cases.}
\label{fig:explanation}
\end{figure}

We make the surprising observation that the modulation signal is extremely sensitive to the assumed properties of the Dirac material. For the case of ZrTe$_5$ both the total rate and the modulation amplitude differ substantially for the different values of the Fermi velocities and the band gap. This is investigated more closely in Fig.~\ref{fig:explanation}, which in the left panel shows the derivative of the total rate with respect to the cosine of the angle $\psi$ between the momentum transfer $\mathbf{q}$ and the velocity of the Earth $\mathbf{v}_\mathrm{e}$. We can see that for $t = 0.5 \, \text{days}$ (i.e.\ close to the maximum of the rate) the differential event rate looks similar in the two cases and is strongly peaked towards $\cos \psi = 1$, such that the momentum transfer is aligned with the direction of the DM wind.

For $t = 0 \, \text{days}$ on the other hand, there are decisive differences between the two cases. While for the theoretically calculated Fermi velocities and band gap the differential rate still peaks at $\cos \psi = 1$, for the experimental values scattering with $\cos \psi \approx 1$ is strongly suppressed. This can be traced back to the fact that in this case the Fermi velocity pointing in the direction of the DM wind is $v_{\mathrm{F}} = 1.6 \cdot 10^{-3}$ and hence close to the maximum velocity of DM particles in the Galactic halo. As a result, only very few DM particles have sufficient kinetic energy to induce scattering with $\cos \psi \approx 1$ and the event rate is suppressed.

As a result, for the theoretically calculated properties of ZrTe$_5$ we find a larger total rate but a smaller modulation amplitude than for the experimentally measured properties. This is illustrated in the right panel in Fig.~\ref{fig:explanation}, which shows the modulation amplitude (blue) and the significance for a daily modulation for $\sigma_e = 10^{-41} \, \mathrm{cm^2}$ (orange) in the two cases. We can see that for the experimentally measured properties the modulation amplitude is substantially larger and hence the significance of a daily modulation is increased in spite of the smaller total rate.

Finally, we note that for the theoretical properties of ZrTe$_5$ the amplitude of the modulation vanishes for $m_\chi \sim 500 \, \mathrm{keV}$ and becomes negative for larger DM masses. This is a result of two competing effects: The velocity integral gives the largest contribution if the DM wind points in the direction of the smallest Fermi velocity. At the same time, the combination of form factors $F_{\text{DM}}$ and $f_{\mathbf{k}\rightarrow\mathbf{k}^\prime}$ favors small $\mathbf{q}$ but large $\tilde{\mathbf{q}}$. It, hence, prefers scattering in the direction of the larger Fermi velocities. For small DM masses, the former effect dominates and leads to a modulation peaked at $t = 0.5 \, \mathrm{days}$ while for larger DM masses the second effect can be comparable or even dominant.\footnote{This is because the minimal velocity for scattering $v_{\text{min}}$ decreases with mass (cf.~\ref{eq:vmininequality}) such that the suppression of the velocity integral in the direction of the large Fermi velocity becomes less pronounced.}
This can lead to a vanishing modulation amplitude for specific values of the DM mass or even an anti-modulation peaked at $t = 0 \, \mathrm{days}$. Since our definition of $\chi^2$ is symmetric in $N_\text{max}$ and $N_\text{min}$ the case of anti-modulation is automatically included in our test for daily modulation. 

An interesting side remark concerns the dark matter form factor. Since we focused on the exchange of a light dark photon mediator, the latter was taken to scale as $F_{\text{DM}}\propto q^{-2}$ with the four-momentum transfer. This behavior changes if we consider heavy mediator exchange for which $F_{\text{DM}}$ approaches a constant. We find that the momentum scaling of $F_{\text{DM}}$ has profound implications on the modulation of the DM scattering rate. For illustration, we depict the sensitivity of Dirac materials for the heavy mediator case in App.~\ref{sec:heavymediator}. Most remarkably, the modulation fraction is increased and the flip in the modulation amplitude at $m_\chi \sim 500 \, \mathrm{keV}$ completely disappears for ZrTe$_5$.

\begin{figure}[t!]
\begin{center}   
 \includegraphics[width=0.4\textwidth]{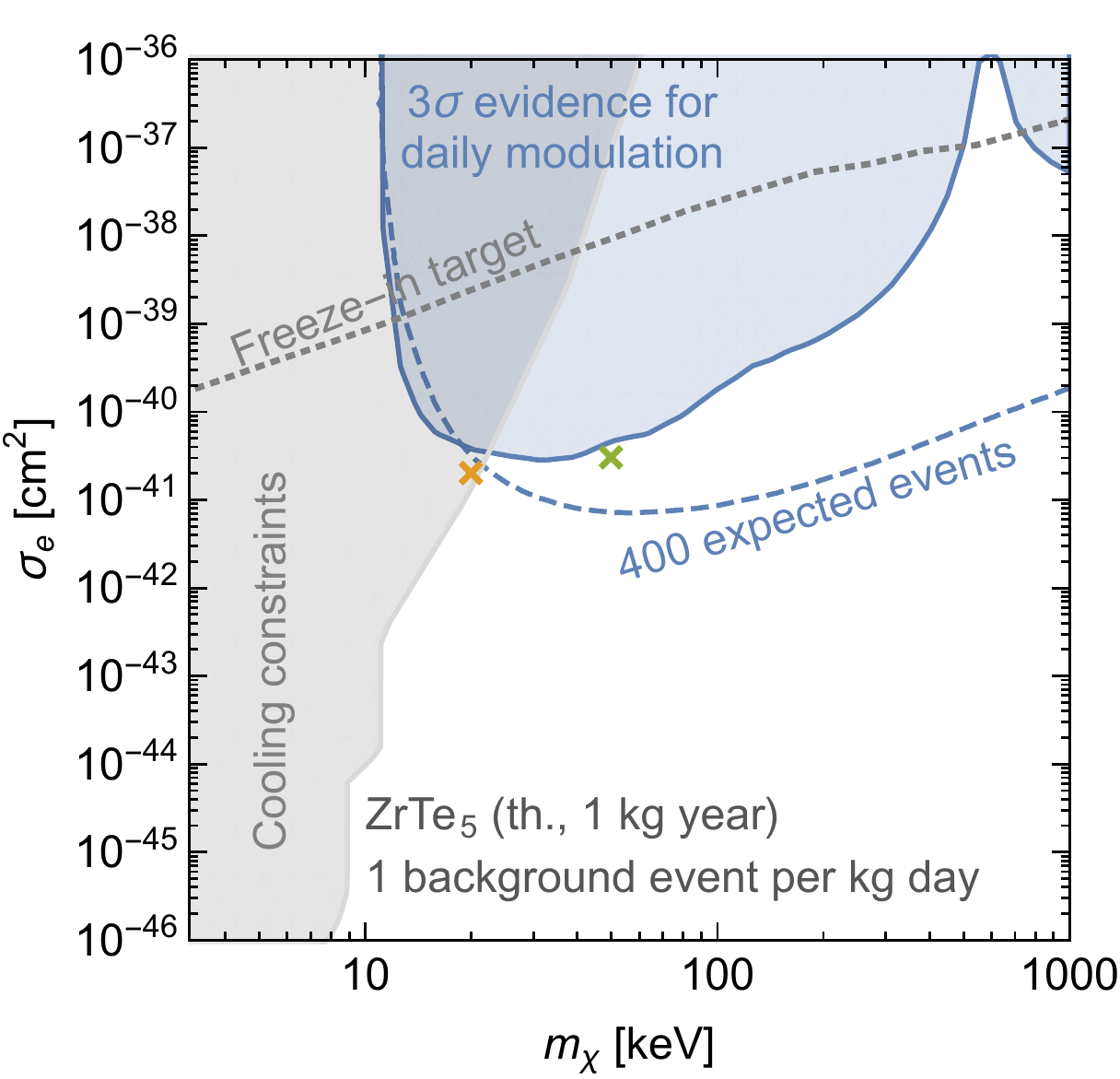}
  \includegraphics[width=0.4\textwidth]{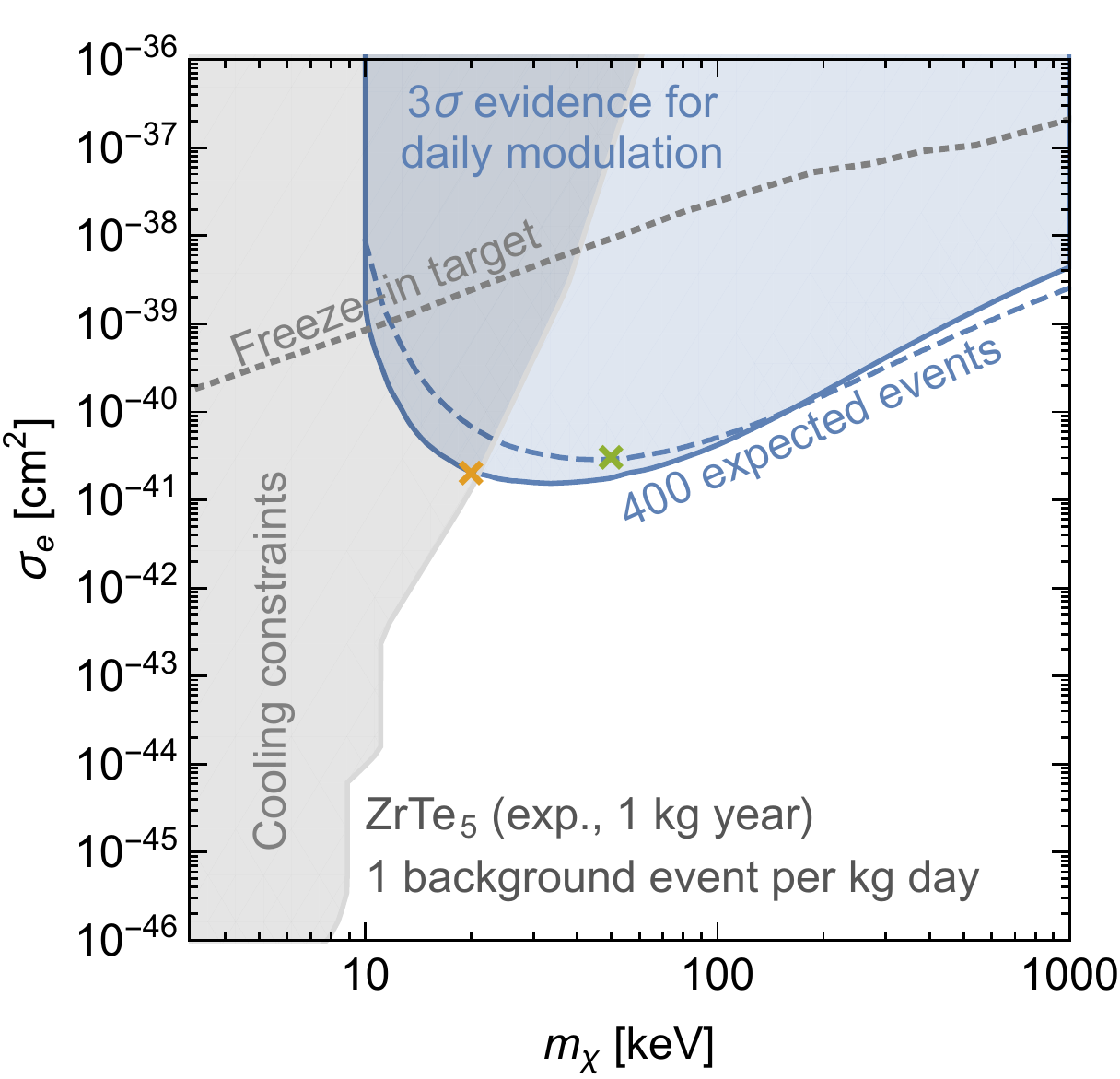}
  
  \includegraphics[width=0.4\textwidth]{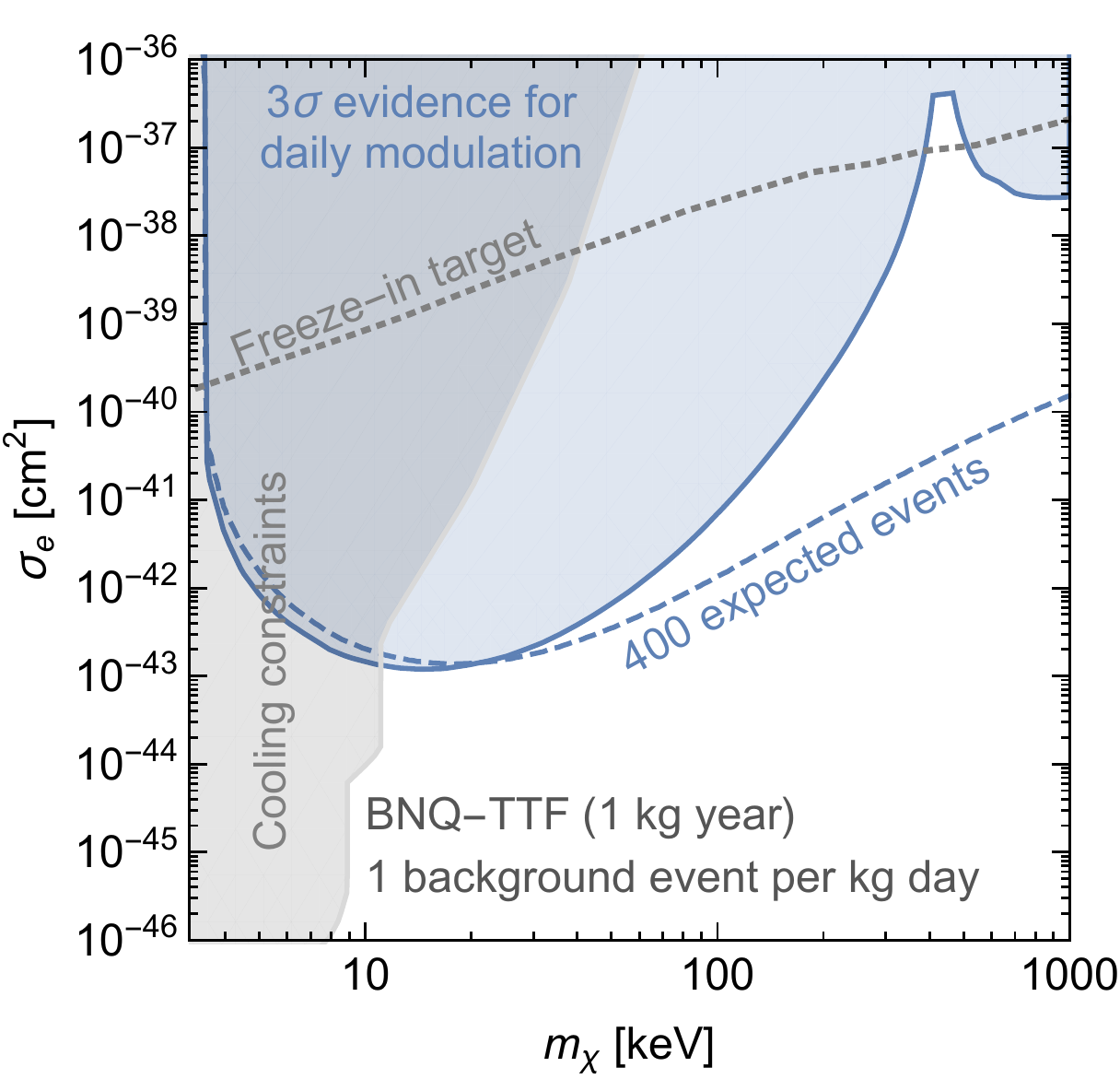}
  \includegraphics[width=0.4\textwidth]{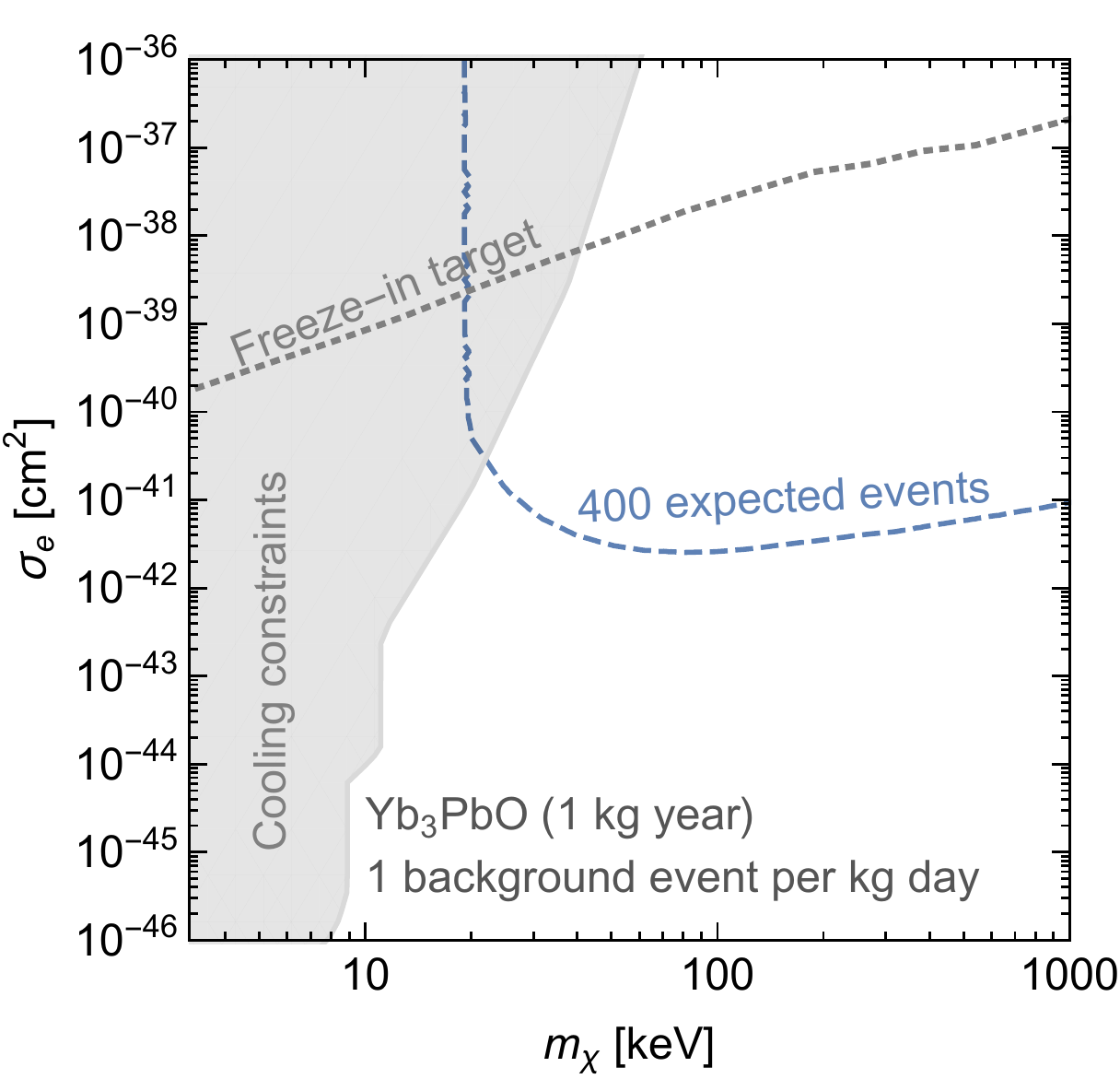}
\end{center}
\caption{Same as Fig.~\ref{fig:results} but under the assumption of a background rate of 1 event per kg day.}
\label{fig:background}
\end{figure}

Thanks to its tiny band gap the organic Dirac material BNQ-TTF can probe significantly smaller DM masses than ZrTe$_5$. For the assumed value $\Delta = 5\,\mathrm{keV}$ the sensitivity extends down to $m_\chi > 4\,\mathrm{keV}$. Close to the threshold the modulation amplitude is found to be quite large, but it decreases rapidly for heavier DM particles and switches sign for $m_\chi > 100\,\mathrm{keV}$. 

For the last material Yb$_3$PbO we only show the sensitivity based on the absolute rate. The modulation signal is suppressed due to the very symmetric nature of this material.

Finally, we consider the case where backgrounds are non-negligible and assume for concreteness a background rate of 1 event per kg day (corresponding to 365 events from background in the assumed exposure of 1 kg year). The estimated sensitivities in this case are shown in Fig.~\ref{fig:background}. In this case, only parameter points predicting more than 400 signal events can be excluded based on the absolute rate, and the resulting bounds are therefore much weaker than in Fig.~\ref{fig:results}. However, the parameter region where a DM signal can be identified based on its daily modulation remains almost unchanged. In fact, for DM masses close to the kinematic threshold, the modulation fraction can be so large that a DM signal can be identified even if the DM signal is significantly smaller than the number of background events. 

\begin{figure}[t]
\begin{center}   
 \includegraphics[height=6cm]{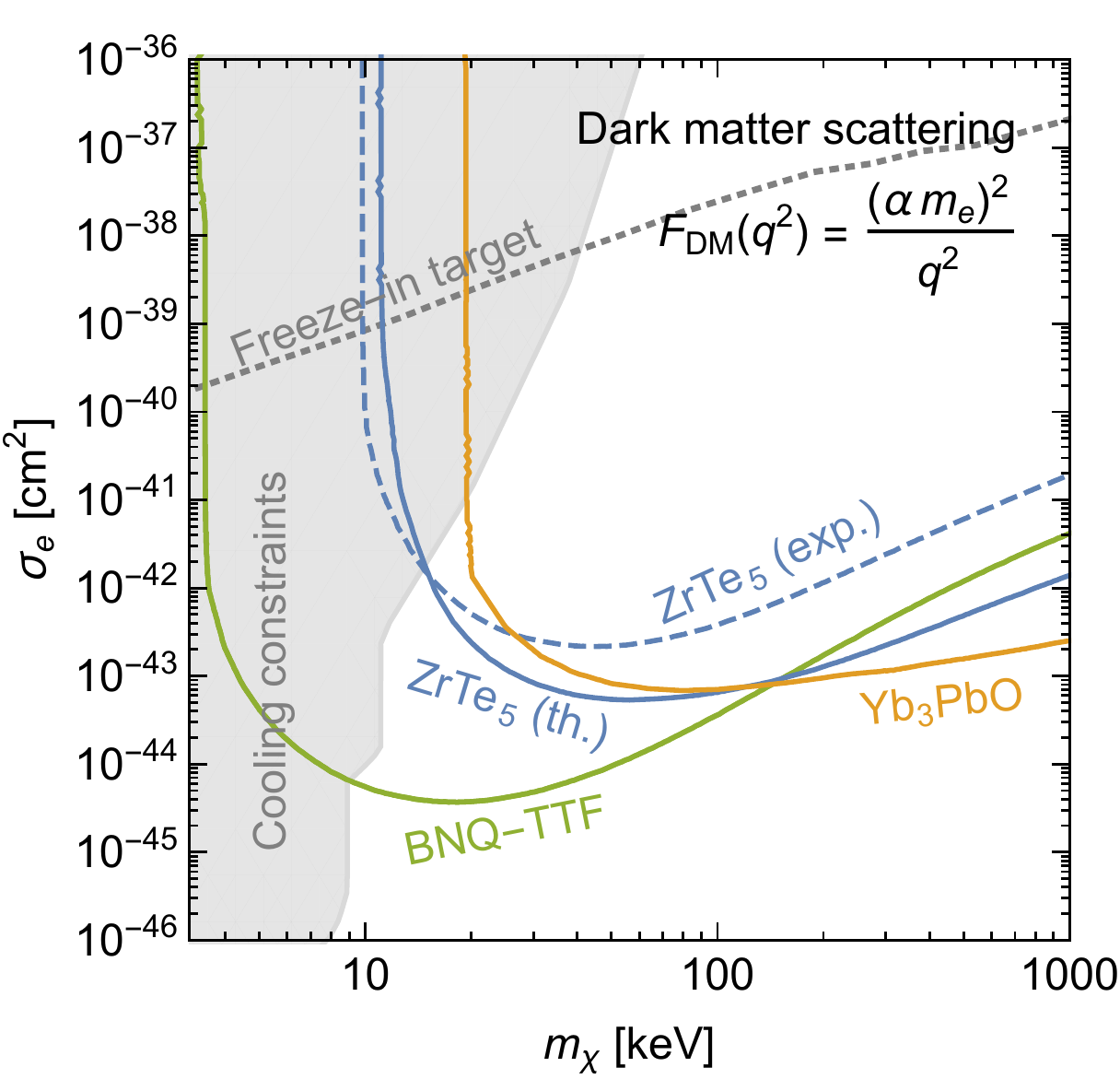}\qquad
  \includegraphics[height=6cm]{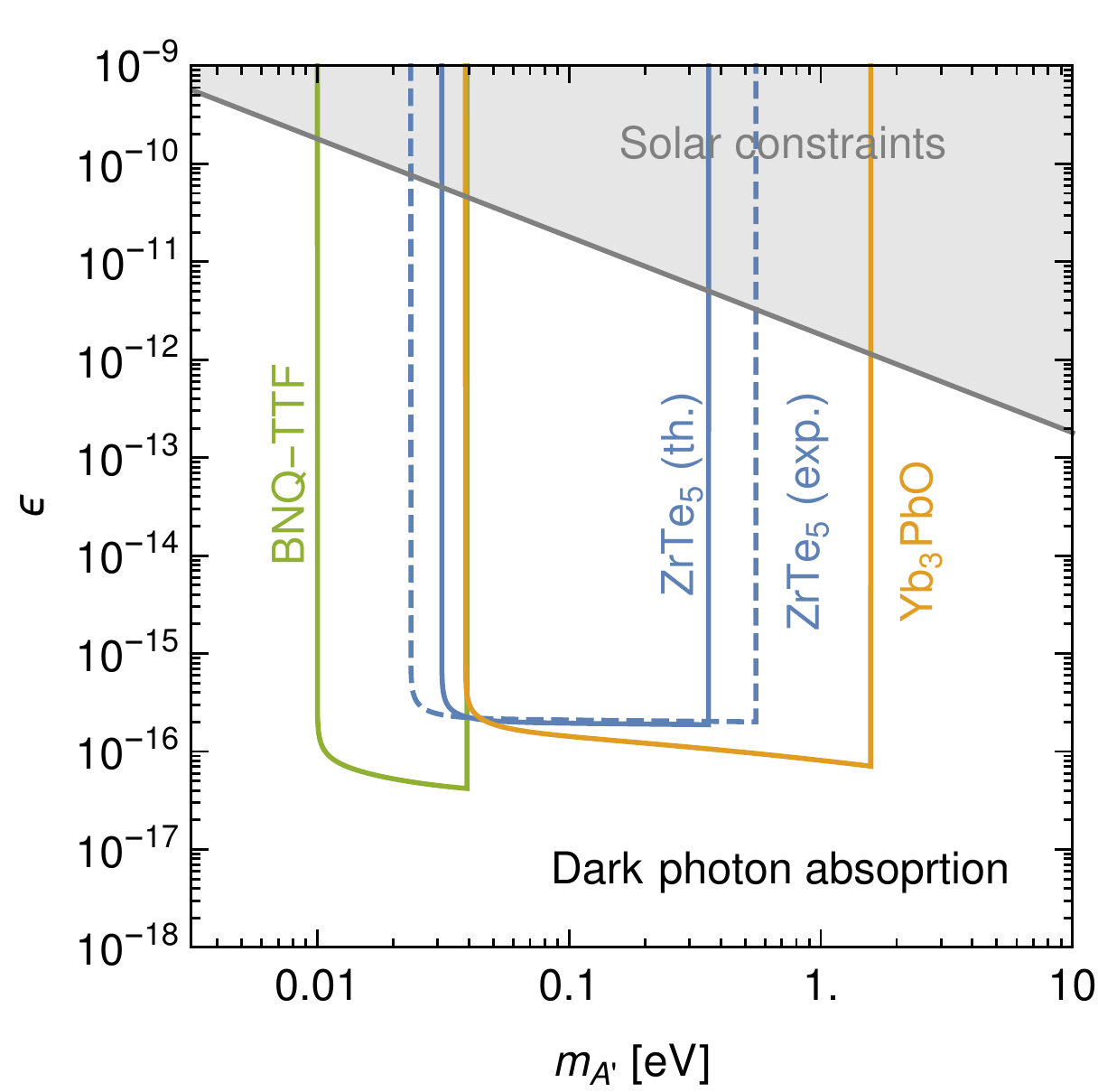}  
\end{center}
\caption{Experimental sensitivity based on the total rate and under the assumption of no backgrounds for the case of DM scattering (left) and dark photon absorption (right). For all experiments we have assumed an exposure of $1\,\mathrm{kg \, year}$. Also shown are astrophysical constraints from red giant and white dwarf cooling~\cite{Vogel:2013raa} (left) and from solar dark photon emission~\cite{An:2013yfc,Vinyoles:2015aba} (right).}
\label{fig:summaryplots}
\end{figure}

To conclude this discussion, let us briefly comment on the dependence of our results on the cut-off $\Lambda$. It is clear that introducing such a cut-off leads to conservative results for the total rate, since only a part of the Brillouin zone is included in the calculation. However, for the modulation amplitude it could in principle happen that the region excluded from the calculation contributes to the modulation with the opposite phase and would hence reduce rather than increase the modulation amplitude. We have therefore explicitly confirmed that variations in the cut-off $\Lambda$ do not significantly modify any of the results presented in this section. The reason is that the dominant contribution to DM scattering stems from collisions with small momentum transfer, for which the precise value of the cut-off is irrelevant. Only if the modulation amplitude nearly vanishes (i.e.\ close to the transition from modulation to anti-modulation) can the dependence on $\Lambda$ be sizeable. Since the search for a daily modulation is essentially insensitive in this particular region, this dependence on $\Lambda$ is of little practical importance.

\subsection{Summary for Dark Matter Scattering and Absorption}

Our sensitivity studies for the three considered Dirac materials are summarized in Fig~\ref{fig:summaryplots}. The left panel covers DM scattering, while the right panel refers to dark photon absorption. Intriguingly, the two new materials suggested in this work, BNQ-TTF and Yb$_3$PbO, reach a very competitive sensitivity compared to ZrTe$_5$ for both cases. The smallness of the band gap and Fermi velocities make BNQ-TTF the ideal target to search for scattering (absorption) of DM particles with masses down to a few keV (meV). Yb$_3$PbO, on the other hand, has a relatively large band gap and is therefore only sensitive to $m_\chi \gtrsim 20\,\mathrm{keV}$ and $m_{A^\prime} \gtrsim 40\,\mathrm{meV}$ respectively. While the large amount of symmetry makes this material unsuitable to search for a daily modulation, the small Fermi velocities combined with the large cutoff scale imply the best sensitivity to DM particles with $m_\chi > 100 \,\mathrm{keV}$ ($m_{A^\prime} \gtrsim 50\,\mathrm{meV}$) based on the total rate alone.

The comparison of the projected sensitivities with existing constraints is quite striking. For the case of DM-electron scattering, DM masses below about $10\,\mathrm{keV}$ are robustly excluded by considerations of stellar cooling in white dwarfs and red giants~\cite{Vogel:2013raa}. For larger DM masses, on the other hand, the leading constraint comes from SN1987A, which is compatible with $\sigma_e \lesssim 10^{-35}\,\mathrm{cm}^2$~\cite{Chang:2018rso}. Dirac materials may therefore improve on these constraints by up to eight orders of magnitude. A similar picture emerges for dark photon absorption, where Dirac material detectors could improve existing limits on the kinetic mixing by $4-6$ orders of magnitude in the range $m_{A^\prime}=10\:\text{meV}-1\:\text{eV}$. In order to highlight this impressive sensitivity, let us note that even the extremely tiny kinetic mixing induced by gravity at six loop order~\cite{Gherghetta:2019coi} which has $\epsilon \lesssim 10^{-13}$ is within reach for Dirac materials.

\section{Conclusions}
\label{sec:summary}

Detectors built from Dirac materials with sub-eV band gap are one of the most promising strategies to search for sub-MeV DM particles interacting with electrons via the exchange of a dark photon. At the same time they can search directly for the absorption of dark photons with sub-eV masses. In the present work we have studied the properties of several different Dirac materials in order to answer the question how a potential DM signal in such a material can be distinguished from backgrounds. The central observation is that in anisotropic materials the DM signal is predicted to exhibit a daily modulation due to the rotation of the Earth relative to the incoming DM wind, which can be used to reject the background hypothesis.

In the first part of this work (Secs.~\ref{sec:formalism} and~\ref{sec:polarizationtensor}) we have revisited the formalism to calculate experimental event rates for the scattering or absorption of DM particles in anisotropic Dirac materials and provided a number of improvements to previous results:
\begin{itemize}
 \item In eq.~(\ref{eq:gtilde}) we introduce a simple way to include the anisotropy of the DM velocity distribution by calculating the velocity integral in terms of the minimum velocity $v_\text{min}$ and the angle $\psi$ between the momentum transfer and the velocity of the Earth.
 \item In eq.~(\ref{eq:Lambdatilde}) we propose an improved way of defining the cut-off $\tilde{\Lambda}$ that determines the region of reciprocal space where the electrons behave like a free Dirac fermion.
 \item We show that in the case of (unpolarized) dark photon absorption there is no daily modulation. 
 This conclusion is drawn from the fact that the absorption rate is determined by the spatial components of the polarization tensor, see eq.~(\ref{eq:absorption}). The latter is found to be independent of the three-momentum transfer in the kinematic regime relevant for absorption ($\omega\gg |\mathbf{q}|$), see eq.~(\ref{eq:tensor2}).
 \item Eq.~(\ref{eq:Fmed}) provides the correct expression for the in-medium form factor $\mathcal{F}_\text{med}$ in the case that the dielectric tensor is anisotropic.
 \item We point out that the daily modulation depends sensitively on the assumed DM form factor and that the modulation amplitude is significantly larger for the case of a heavy mediator than for a light mediator (see App.~\ref{sec:heavymediator}).
\end{itemize}

In the second part of this work (Sec.~\ref{sec:candidates}) we have presented a number of candidate Dirac materials that possess the required properties to detect DM particles in the sub-MeV range. We have performed an improved calculation of the band structure of ZrTe$_5$ and determined the band gap, Fermi velocities and the dielectric tensor (see Tabs.~\ref{diracvelocities} and~\ref{dielectric}). In particular, we have confirmed the finding that this material exhibits a sizeable anisotropy, which makes it particularly well-suited to search for a daily modulation.

We furthermore propose two new Dirac materials, BNQ-TTF and Yb$_3$PbO, which have not been previously considered in the context of DM physics. Both materials have significantly smaller Fermi velocities and therefore potentially much larger sensitivity to DM scattering than ZrTe$_5$. While Yb$_3$PbO crystallizes in a cubic lattice and therefore exhibits little anisotropy, BNQ-TTF is found to be highly anisotropic and furthermore exhibits a tiny band gap, making this material extremely attractive for further investigations. As reported in Ref. \cite{BNQ}, a macroscopic sample of BNQ-TTF has already been synthesized, feasible for usage in devices. It will be exciting to see whether the properties that we predict can be confirmed in the laboratory.

Finally, in Sec.~\ref{sec:results} we provided our sensitivity estimates for the three Dirac materials (see Fig.~\ref{fig:results}). We have identified the parameter regions that can be excluded by a null result as well as the parameter regions where the daily modulation is large enough to provide a way to confirm the DM nature of an observed signal. The statistical method that we use to search for daily modulations can easily be extended to include a non-modulating background contribution and we find that anisotropic Dirac materials retain an impressive sensitivity to DM scattering even in the presence of sizeable backgrounds (see Fig.~\ref{fig:background}). However, we also conclude that the modulation signal depends sensitively on the properties of the Dirac material, in particular the Fermi velocities, making a precise determination of these properties essential.

Clearly, there is still a long way to go before the first DM detector based on a Dirac material will be built. Nevertheless, as the interest for DM models in the sub-MeV mass range grows rapidly, there will be an increasing incentive to exploit the great potential of this technology. Both improved calculations and experimental measurements will be essential in order to identify the materials most suited for exploring this uncharted territory of DM physics.

\section*{Acknowledgement}\label{acknowledgements}

We thank Alexander Balatsky, Riccardo Catena, Jan Conrad, Timon Emken, Alfredo Ferella, Katherine Freese, Bart Olsthoorn, and Annika Reinert for helpful discussions. FK is funded  by  the  Deutsche Forschungsgemeinschaft (DFG) through the  Emmy Noether Grant No.\ KA 4662/1-1 and the Collaborative Research Center TRR 257 ``Particle Physics Phenomenology after the Higgs Discovery''. RMG acknowledges funding from the VILLUM FONDEN via the Centre of Excellence for Dirac Materials (Grant No. 11744) and the European Research Council ERC HERO grant. MWW acknowledges support by the Vetenskapsr\r{a}det (Swedish Research Council) through contract No. 638-2013-8993 and the Oskar Klein Centre for Cosmoparticle Physics. The authors acknowledge computational resources from the Swedish National Infrastructure for Computing (SNIC) at the National Supercomputer Centre at Linköping University, the Centre for High Performance Computing (PDC), the High Performance Computing Centre North (HPC2N), and the Uppsala Multidisciplinary Centre for Advanced Computational Science (UPPMAX).

\section*{Note Added}\label{acknowledgements}

While this work was nearing completion, Ref.~\cite{Coskuner:2019odd} appeared, which also considers the daily modulation of DM signals in Dirac Materials. For the case of DM scattering in ZrTe$_5$ our results are in qualitative agreement, but there are important differences for the case of DM absorption, as discussed in detail in the text.

\appendix
\section{Sensitivity of Dirac Materials for DM Scattering with Heavy Mediators}\label{sec:heavymediator}

\begin{figure}[t!]
\begin{center}   
 \includegraphics[width=0.4\textwidth]{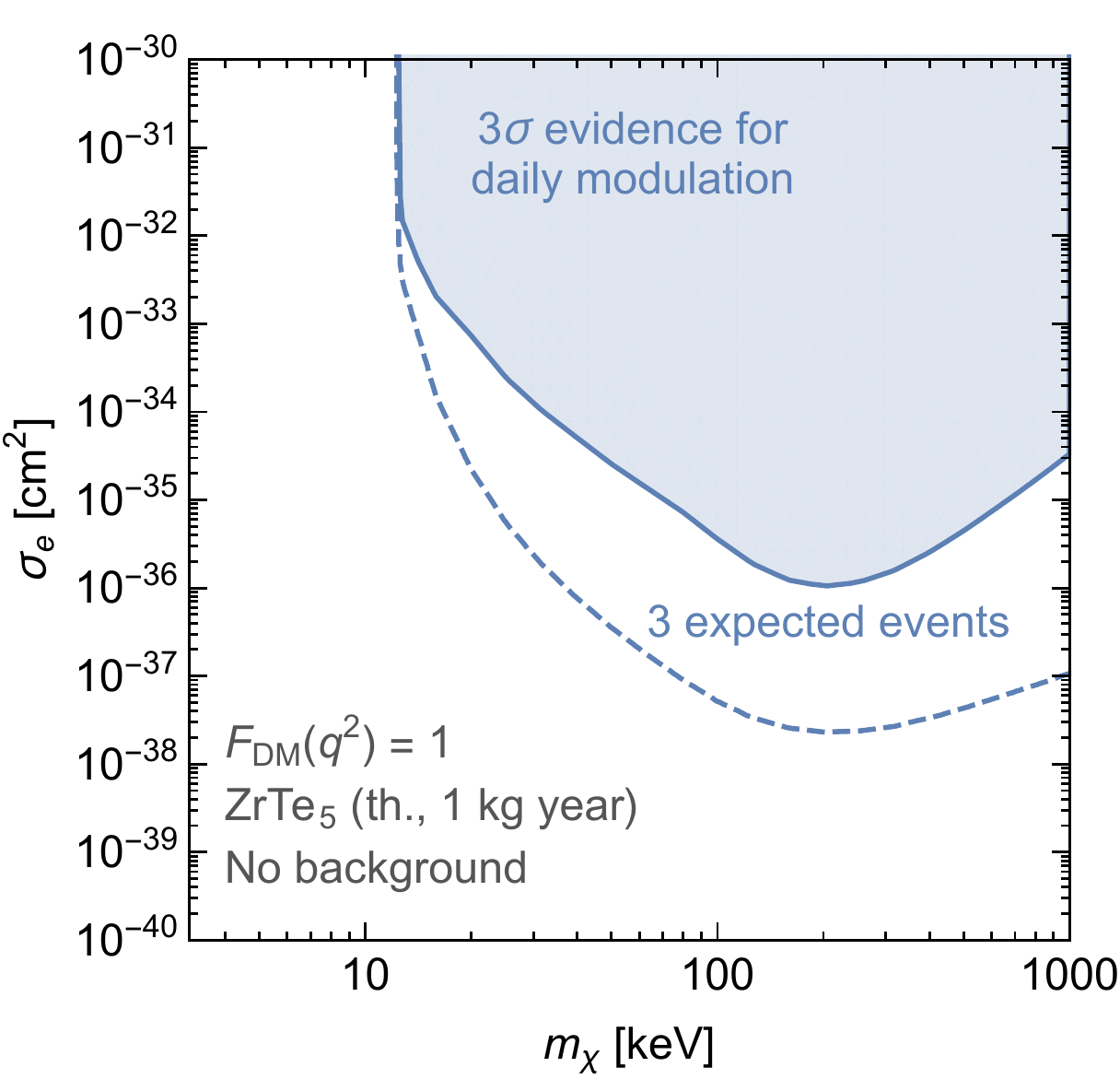}
  \includegraphics[width=0.4\textwidth]{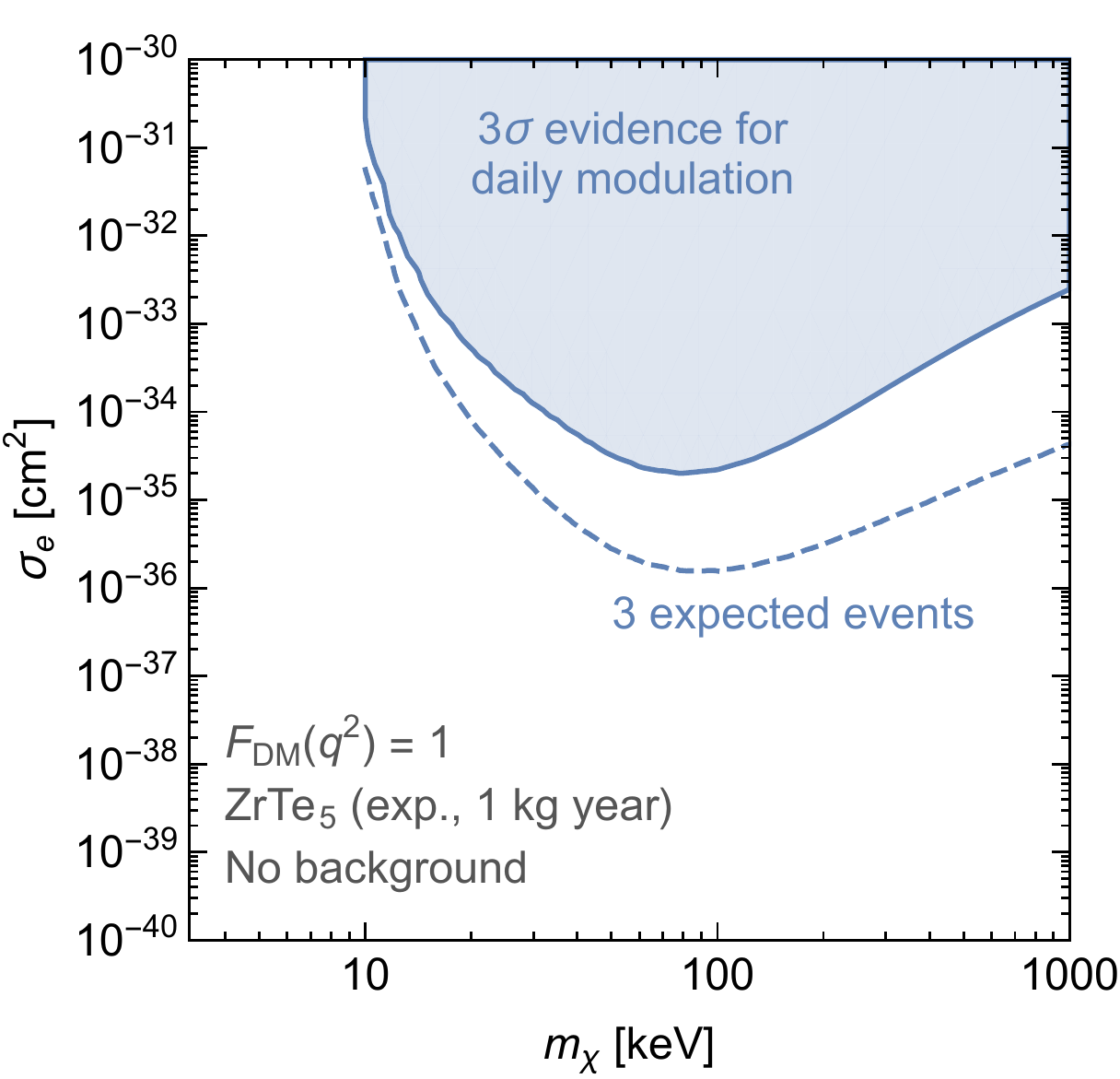}
  
  \includegraphics[width=0.4\textwidth]{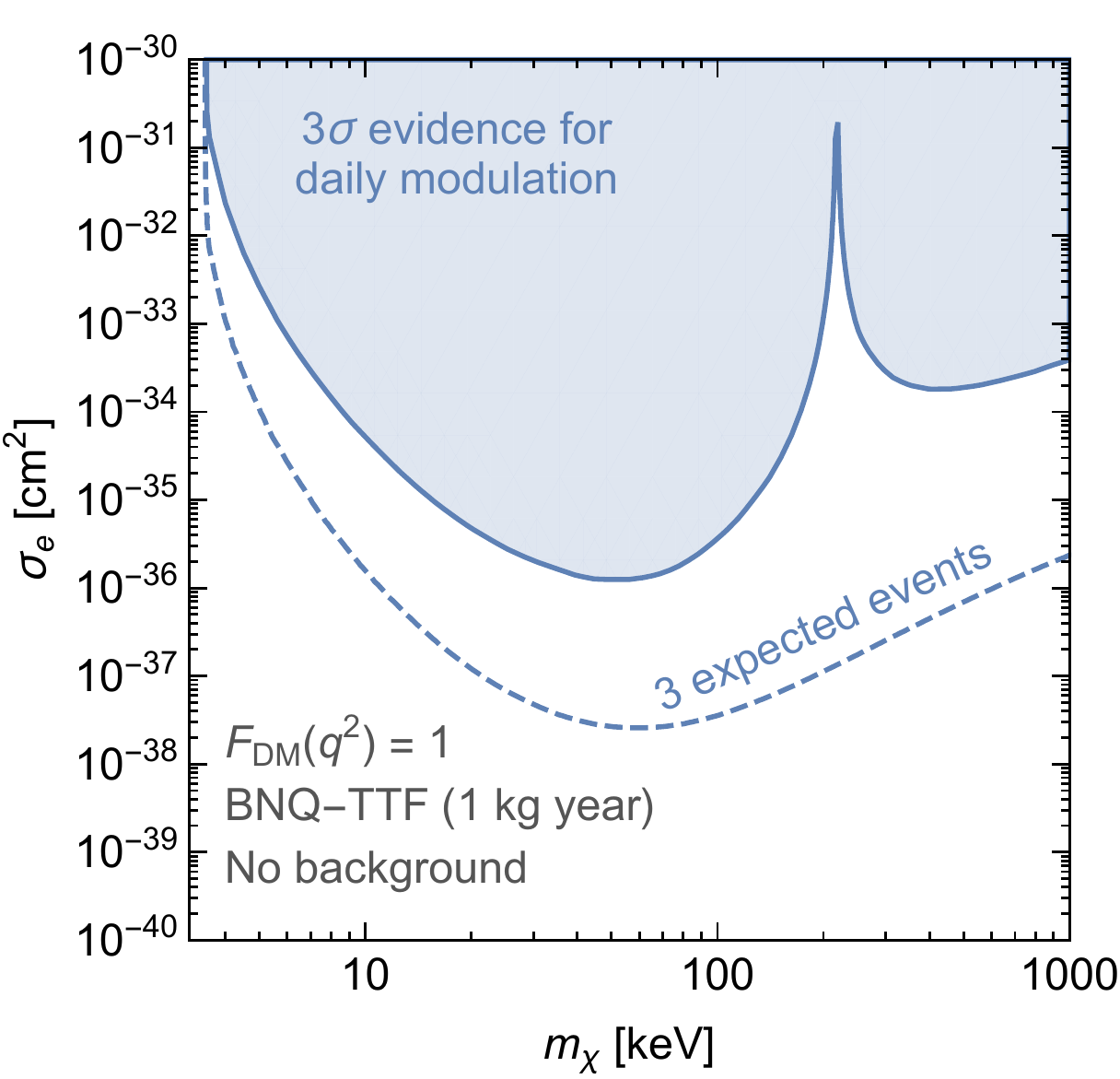}
  \includegraphics[width=0.4\textwidth]{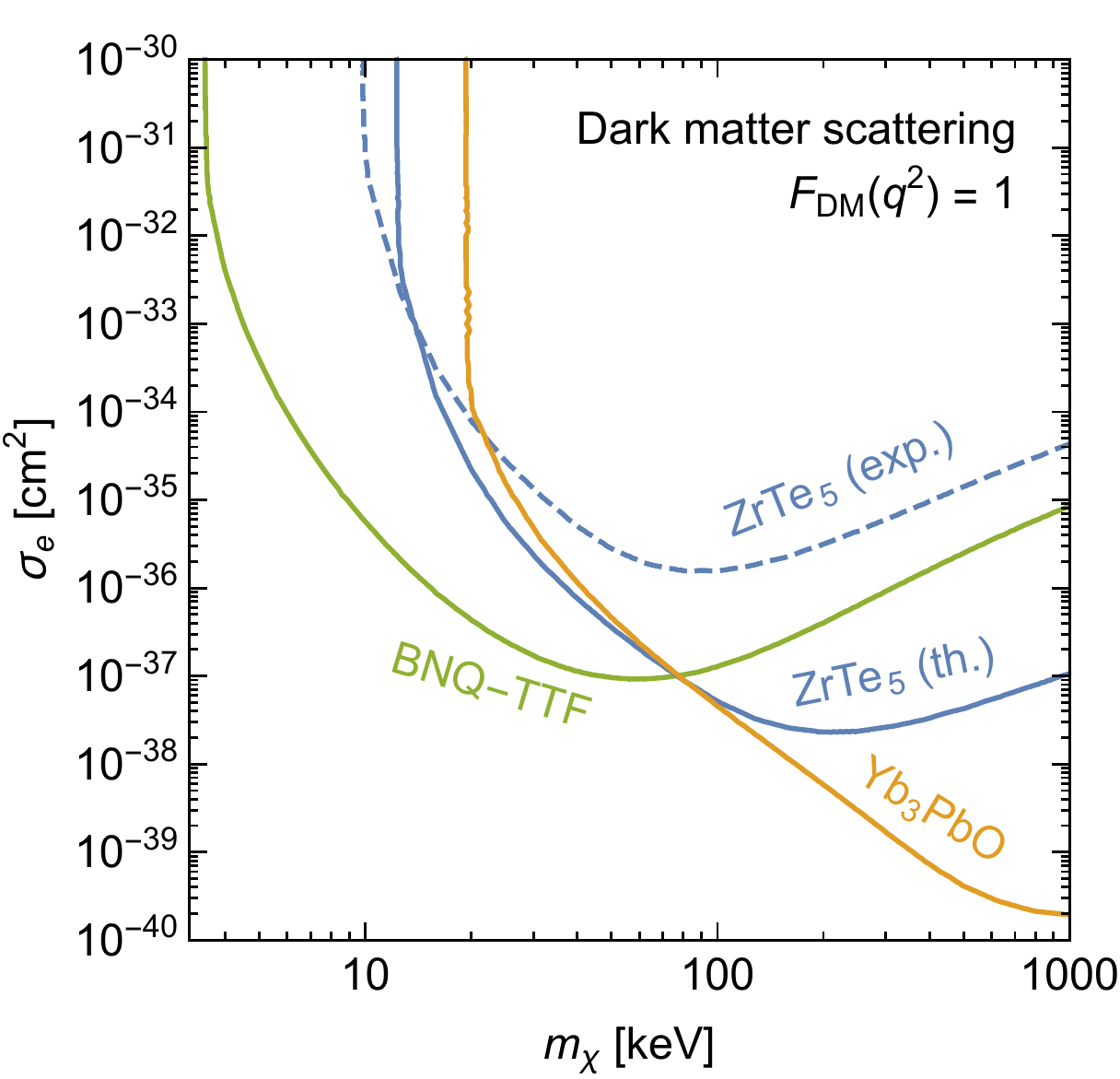}

\end{center}
\caption{Same as Fig.~\ref{fig:results} but for the case of a heavy mediator ($F_\text{DM} = 1$). In the final panel we compare the experimental sensitivity based on the total rate for all materials.}
\label{fig:results_contact}
\end{figure}

In this appendix we consider the case of a heavy dark photon with $m_{A'} \gg |\mathbf{q}|$. In this case the DM-electron scattering cross section becomes independent of the momentum transfer and the DM form factor simply becomes $F_\text{DM}(q) = 1$. Compared to the case of a light dark photon, this leads to a strong suppression of the scattering rate for small DM masses (i.e. small momentum transfer) but a much milder suppression for large DM masses (large momentum transfer). This can be seen in Fig.~\ref{fig:results_contact}, which shows the parameter regions that could be excluded by an experimental null result (corresponding to 3 expected events) as well as the parameter regions where one could detect a daily modulation at $3\sigma$ significance.

We observe that for the case of a heavy mediator the modulation fraction is generally enhanced. This is because the DM form factor for a heavy mediator leads to a much weaker preference for scattering in the direction of large Fermi velocities than the DM form factor for a light mediator. For example, in ZrTe$_5$ we find the modulation fraction to be greater than $10\%$ for all DM masses and no change of sign for large DM masses. Unfortunately, we find that for large DM masses our sensitivity estimates depend on the adopted value of the cut-off $\Lambda$, because scattering with momenta close to the cut-off gives an important contribution. This explains in particular why the sensitivity of ZrTe$_5$ is much worse when using the properties from Ref.~\cite{hochberg} (including $\Lambda = 0.07 \,$\AA) than for the properties determined from our calculations (for which $\Lambda$ is significantly larger). The sensitivities shown in Fig.~\ref{fig:results_contact} should therefore be interpreted as conservative estimates.\footnote{We have checked explicitly that our estimate of the modulation amplitude is also conservative, i.e.\ increasing the cut-off $\Lambda$ would lead to a larger modulation amplitude.}

Compared to $m_{A^\prime}=0$, the heavy mediator case is more strongly constrained by cosmological and astrophysical observations like the effective neutrino number and supernovae (see e.g.\ Ref.~\cite{Chang:2019xva}). However, in the heavy mediator case, the cosmological bounds on $\bar{\sigma}_e$ are sensitive to the choice of the dark sector gauge coupling and the precise dark photon mass. Since heavy mediators are not within our main focus, a more detailed investigation of the corresponding astrophysical and cosmological constraints is beyond the scope of this work.

\providecommand{\href}[2]{#2}\begingroup\raggedright\endgroup


\begin{thebibliography}{10}

\bibitem{Athron:2018hpc}
{\bf GAMBIT Collaboration}, P.~Athron et~al.,
  \href{http://dx.doi.org/10.1140/epjc/s10052-018-6513-6}{{\it {Global analyses
  of Higgs portal singlet dark matter models using GAMBIT}}, } {\em Eur. Phys.
  J.} {\bf C79} (2019), no.~1 38, [\href{http://arxiv.org/abs/1808.10465}{{\tt
  1808.10465}}].

\bibitem{Knapen:2017xzo}
S.~Knapen, T.~Lin, and K.~M. Zurek,
  \href{http://dx.doi.org/10.1103/PhysRevD.96.115021}{{\it {Light Dark Matter:
  Models and Constraints}}, } {\em Phys. Rev.} {\bf D96} (2017), no.~11 115021,
  [\href{http://arxiv.org/abs/1709.07882}{{\tt 1709.07882}}].

\bibitem{Lin:2019uvt}
T.~Lin, \href{http://dx.doi.org/10.22323/1.333.0009}{{\it {Dark matter models
  and direct detection}}, } {\em PoS} {\bf 333} (2019) 009,
  [\href{http://arxiv.org/abs/1904.07915}{{\tt 1904.07915}}].

\bibitem{Hall:2009bx}
L.~J. Hall, K.~Jedamzik, J.~March-Russell, and S.~M. West,
  \href{http://dx.doi.org/10.1007/JHEP03(2010)080}{{\it {Freeze-In Production
  of FIMP Dark Matter}}, } {\em JHEP} {\bf 03} (2010) 080,
  [\href{http://arxiv.org/abs/0911.1120}{{\tt 0911.1120}}].

\bibitem{Essig:2011nj}
R.~Essig, J.~Mardon, and T.~Volansky,
  \href{http://dx.doi.org/10.1103/PhysRevD.85.076007}{{\it {Direct Detection of
  Sub-GeV Dark Matter}}, } {\em Phys. Rev.} {\bf D85} (2012) 076007,
  [\href{http://arxiv.org/abs/1108.5383}{{\tt 1108.5383}}].

\bibitem{Chu:2011be}
X.~Chu, T.~Hambye, and M.~H.~G. Tytgat,
  \href{http://dx.doi.org/10.1088/1475-7516/2012/05/034}{{\it {The Four Basic
  Ways of Creating Dark Matter Through a Portal}}, } {\em JCAP} {\bf 1205}
  (2012) 034, [\href{http://arxiv.org/abs/1112.0493}{{\tt 1112.0493}}].

\bibitem{Green:2017ybv}
D.~Green and S.~Rajendran,
  \href{http://dx.doi.org/10.1007/JHEP10(2017)013}{{\it {The Cosmology of
  Sub-MeV Dark Matter}}, } {\em JHEP} {\bf 10} (2017) 013,
  [\href{http://arxiv.org/abs/1701.08750}{{\tt 1701.08750}}].

\bibitem{Heeba:2018wtf}
S.~Heeba, F.~Kahlhoefer, and P.~Stöcker,
  \href{http://dx.doi.org/10.1088/1475-7516/2018/11/048}{{\it {Freeze-in
  production of decaying dark matter in five steps}}, } {\em JCAP} {\bf 1811}
  (2018), no.~11 048, [\href{http://arxiv.org/abs/1809.04849}{{\tt
  1809.04849}}].

\bibitem{An:2018nvz}
H.~An, R.~Huo, and W.~Liu, {\it {KeV Scale Frozen-in Self-Interacting Fermionic
  Dark Matter}},  \href{http://arxiv.org/abs/1812.05699}{{\tt 1812.05699}}.

\bibitem{Dvorkin:2019zdi}
C.~Dvorkin, T.~Lin, and K.~Schutz,
  \href{http://dx.doi.org/10.1103/PhysRevD.99.115009}{{\it {Making dark matter
  out of light: freeze-in from plasma effects}}, } {\em Phys. Rev.} {\bf D99}
  (2019), no.~11 115009, [\href{http://arxiv.org/abs/1902.08623}{{\tt
  1902.08623}}].

\bibitem{Hochberg:2015pha}
Y.~Hochberg, Y.~Zhao, and K.~M. Zurek,
  \href{http://dx.doi.org/10.1103/PhysRevLett.116.011301}{{\it {Superconducting
  Detectors for Superlight Dark Matter}}, } {\em Phys. Rev. Lett.} {\bf 116}
  (2016), no.~1 011301, [\href{http://arxiv.org/abs/1504.07237}{{\tt
  1504.07237}}].

\bibitem{Hochberg:2016ajh}
Y.~Hochberg, T.~Lin, and K.~M. Zurek,
  \href{http://dx.doi.org/10.1103/PhysRevD.94.015019}{{\it {Detecting
  Ultralight Bosonic Dark Matter via Absorption in Superconductors}}, } {\em
  Phys. Rev.} {\bf D94} (2016), no.~1 015019,
  [\href{http://arxiv.org/abs/1604.06800}{{\tt 1604.06800}}].

\bibitem{Hochberg:2019cyy}
Y.~Hochberg, I.~Charaev, S.-W. Nam, V.~Verma, M.~Colangelo, et~al., {\it
  {Detecting Dark Matter with Superconducting Nanowires}},
  \href{http://arxiv.org/abs/1903.05101}{{\tt 1903.05101}}.

\bibitem{Schutz:2016tid}
K.~Schutz and K.~M. Zurek,
  \href{http://dx.doi.org/10.1103/PhysRevLett.117.121302}{{\it {Detectability
  of Light Dark Matter with Superfluid Helium}}, } {\em Phys. Rev. Lett.} {\bf
  117} (2016), no.~12 121302, [\href{http://arxiv.org/abs/1604.08206}{{\tt
  1604.08206}}].

\bibitem{Knapen:2016cue}
S.~Knapen, T.~Lin, and K.~M. Zurek,
  \href{http://dx.doi.org/10.1103/PhysRevD.95.056019}{{\it {Light Dark Matter
  in Superfluid Helium: Detection with Multi-excitation Production}}, } {\em
  Phys. Rev.} {\bf D95} (2017), no.~5 056019,
  [\href{http://arxiv.org/abs/1611.06228}{{\tt 1611.06228}}].

\bibitem{Caputo:2019cyg}
A.~Caputo, A.~Esposito, and A.~D. Polosa, {\it {Sub-MeV Dark Matter and the
  Goldstone Modes of Superfluid Helium}},
  \href{http://arxiv.org/abs/1907.10635}{{\tt 1907.10635}}.

\bibitem{Knapen:2017ekk}
S.~Knapen, T.~Lin, M.~Pyle, and K.~M. Zurek,
  \href{http://dx.doi.org/10.1016/j.physletb.2018.08.064}{{\it {Detection of
  Light Dark Matter With Optical Phonons in Polar Materials}}, } {\em Phys.
  Lett.} {\bf B785} (2018) 386--390,
  [\href{http://arxiv.org/abs/1712.06598}{{\tt 1712.06598}}].

\bibitem{Griffin:2018bjn}
S.~Griffin, S.~Knapen, T.~Lin, and K.~M. Zurek,
  \href{http://dx.doi.org/10.1103/PhysRevD.98.115034}{{\it {Directional
  Detection of Light Dark Matter with Polar Materials}}, } {\em Phys. Rev.}
  {\bf D98} (2018), no.~11 115034, [\href{http://arxiv.org/abs/1807.10291}{{\tt
  1807.10291}}].

\bibitem{Cox:2019cod}
P.~Cox, T.~Melia, and S.~Rajendran,
  \href{http://dx.doi.org/10.1103/PhysRevD.100.055011}{{\it {Dark matter phonon
  coupling}}, } {\em Phys. Rev.} {\bf D100} (2019), no.~5 055011,
  [\href{http://arxiv.org/abs/1905.05575}{{\tt 1905.05575}}].

\bibitem{vergnioryDM}
M.-A. Sanchez-Martinez, I.~Robredo, A.~Bidauzarraga, A.~Bergara, F.~de~Juan,
  et~al., {\it {Spectral and optical properties of Ag$_3$Au(Se$_2$,Te$_2$) and
  dark matter detection}},  \href{http://arxiv.org/abs/1905.04805}{{\tt
  1905.04805}}.

\bibitem{hochberg}
Y.~Hochberg, Y.~Kahn, M.~Lisanti, K.~M. Zurek, A.~G. Grushin, et~al.,
  \href{http://dx.doi.org/10.1103/PhysRevD.97.015004}{{\it {Detection of
  sub-MeV Dark Matter with Three-Dimensional Dirac Materials}}, } {\em Phys.
  Rev.} {\bf D97} (2018), no.~1 015004,
  [\href{http://arxiv.org/abs/1708.08929}{{\tt 1708.08929}}].

\bibitem{Coskuner:2019odd}
A.~Coskuner, A.~Mitridate, A.~Olivares, and K.~M. Zurek, {\it {Directional Dark
  Matter Detection in Anisotropic Dirac Materials}},
  \href{http://arxiv.org/abs/1909.09170}{{\tt 1909.09170}}.

\bibitem{geilhufe2018materials}
R.~M. Geilhufe, B.~Olsthoorn, A.~D. Ferella, T.~Koski, F.~Kahlhoefer, et~al.,
  {\it Materials informatics for dark matter detection},  {\em Physica Status
  Solidi Rapid Research Letters} {\bf 12} (2018), no.~11 1800293.

\bibitem{wehling2014dirac}
T.~Wehling, A.~M. Black-Schaffer, and A.~V. Balatsky, {\it Dirac materials},
  {\em Adv. Phys.} {\bf 63} (2014), no.~1 1--76.

\bibitem{olsthoorn2019mass}
B.~Olsthoorn and A.~V. Balatsky, {\it {Mass fluctuations and absorption rates
  in Dirac materials sensors}},  \href{http://arxiv.org/abs/1909.10394}{{\tt
  1909.10394}}.

\bibitem{dora2009gaps}
B.~Dora and K.~Ziegler, {\it Gaps and tails in graphene and graphane},  {\em
  New Journal of Physics} {\bf 11} (2009), no.~9 095006.

\bibitem{Hochberg:2016ntt}
Y.~Hochberg, Y.~Kahn, M.~Lisanti, C.~G. Tully, and K.~M. Zurek,
  \href{http://dx.doi.org/10.1016/j.physletb.2017.06.051}{{\it {Directional
  detection of dark matter with two-dimensional targets}}, } {\em Phys. Lett.}
  {\bf B772} (2017) 239--246, [\href{http://arxiv.org/abs/1606.08849}{{\tt
  1606.08849}}].

\bibitem{pertsova2019}
A.~Pertsova, R.~M. Geilhufe, M.~Bremholm, and A.~V. Balatsky,
  \href{https://link.aps.org/doi/10.1103/PhysRevB.99.205126}{{\it Computational
  search for dirac and weyl nodes in $f$-electron antiperovskites}, } {\em
  Phys. Rev.} {\bf B99} (May, 2019) 205126.

\bibitem{hofmann2015}
J.~{Hofmann} and S.~{Das Sarma},
  \href{http://dx.doi.org/10.1103/PhysRevB.91.241108}{{\it {Plasmon signature
  in Dirac-Weyl liquids}}, } {\em Phys. Rev.} {\bf B91} (Jun, 2015) 241108,
  [\href{http://arxiv.org/abs/1501.04636}{{\tt 1501.04636}}].

\bibitem{throckmorton}
R.~E. Throckmorton, J.~Hofmann, E.~Barnes, and S.~{Das Sarma},
  \href{http://dx.doi.org/10.1103/PhysRevB.92.115101}{{\it {Many-body effects
  and ultraviolet renormalization in three-dimensional Dirac materials}}, }
  {\em Phys. Rev.} {\bf B92} (2015), no.~11 115101,
  [\href{http://arxiv.org/abs/1505.05154}{{\tt 1505.05154}}].

\bibitem{Essig:2015cda}
R.~Essig, M.~Fernandez-Serra, J.~Mardon, A.~Soto, T.~Volansky, et~al.,
  \href{http://dx.doi.org/10.1007/JHEP05(2016)046}{{\it {Direct Detection of
  sub-GeV Dark Matter with Semiconductor Targets}}, } {\em JHEP} {\bf 05}
  (2016) 046, [\href{http://arxiv.org/abs/1509.01598}{{\tt 1509.01598}}].

\bibitem{Peskin:1995ev}
M.~E. Peskin and D.~V. Schroeder, {\em {An Introduction to quantum field
  theory}}.
\newblock Addison-Wesley, Reading, USA, 1995.

\bibitem{Dyson:1949ha}
F.~J. Dyson, \href{http://dx.doi.org/10.1103/PhysRev.75.1736}{{\it {The S
  matrix in quantum electrodynamics}}, } {\em Phys. Rev.} {\bf 75} (1949)
  1736--1755.

\bibitem{Schwinger:1951ex}
J.~S. Schwinger, \href{http://dx.doi.org/10.1073/pnas.37.7.452}{{\it {On the
  Green's functions of quantized fields. 1.}}, } {\em Proc. Nat. Acad. Sci.}
  {\bf 37} (1951) 452--455.

\bibitem{An:2013yfc}
H.~An, M.~Pospelov, and J.~Pradler,
  \href{http://dx.doi.org/10.1016/j.physletb.2013.07.008}{{\it {New stellar
  constraints on dark photons}}, } {\em Phys. Lett.} {\bf B725} (2013)
  190--195, [\href{http://arxiv.org/abs/1302.3884}{{\tt 1302.3884}}].

\bibitem{An:2013yua}
H.~An, M.~Pospelov, and J.~Pradler,
  \href{http://dx.doi.org/10.1103/PhysRevLett.111.041302}{{\it {Dark Matter
  Detectors as Dark Photon Helioscopes}}, } {\em Phys. Rev. Lett.} {\bf 111}
  (2013) 041302, [\href{http://arxiv.org/abs/1304.3461}{{\tt 1304.3461}}].

\bibitem{thakur}
A.~Thakur, R.~Sachdeva, and A.~Agarwal,
  \href{http://dx.doi.org/10.1088/1361-648X/aa57bd}{{\it {Dynamical
  polarizability, screening and plasmons in one, two and three dimensional
  massive Dirac systems}}, } {\em J. Phys. Condens. Matter} {\bf 29} (Mar,
  2017) 105701, [\href{http://arxiv.org/abs/1604.00806}{{\tt 1604.00806}}].

\bibitem{hamann1979norm}
D.~Hamann, M.~Schl{\"u}ter, and C.~Chiang, {\it Norm-conserving
  pseudopotentials},  {\em Phys. Rev. Lett.} {\bf 43} (1979), no.~20 1494.

\bibitem{blochl1994projector}
P.~E. Bl{\"o}chl, {\it Projector augmented-wave method},  {\em Phys. Rev. B}
  {\bf 50} (1994), no.~24 17953.

\bibitem{pseudo1}
D.~Vanderbilt, {\it Soft self-consistent pseudopotentials in a generalized
  eigenvalue formalism},  {\em Phys. Rev. B} {\bf 41} (1990), no.~11 7892.

\bibitem{pseudo2}
G.~Kresse and J.~Hafner,
  \href{http://stacks.iop.org/0953-8984/6/i=40/a=015}{{\it Norm-conserving and
  ultrasoft pseudopotentials for first-row and transition elements}, } {\em J.
  Phys. Condens. Matter} {\bf 6} (1994), no.~40 8245.

\bibitem{vasp2}
G.~Kresse and J.~Furthm{\"u}ller, {\it Efficient iterative schemes for ab
  initio total-energy calculations using a plane-wave basis set},  {\em Phys.
  Rev. B} {\bf 54} (1996), no.~16 11169.

\bibitem{kresse1999ultrasoft}
G.~Kresse and D.~Joubert, {\it From ultrasoft pseudopotentials to the projector
  augmented-wave method},  {\em Phys. Rev. B} {\bf 59} (1999), no.~3 1758.

\bibitem{sun2015}
J.~Sun, A.~Ruzsinszky, and J.~P. Perdew,
  \href{https://link.aps.org/doi/10.1103/PhysRevLett.115.036402}{{\it Strongly
  constrained and appropriately normed semilocal density functional}, } {\em
  Phys. Rev. Lett.} {\bf 115} (Jul, 2015) 036402.

\bibitem{sun2016accurate}
J.~Sun, R.~C. Remsing, Y.~Zhang, Z.~Sun, A.~Ruzsinszky, et~al., {\it Accurate
  first-principles structures and energies of diversely bonded systems from an
  efficient density functional},  {\em Nature chemistry} {\bf 8} (2016), no.~9
  831.

\bibitem{tkatchenko2009accurate}
A.~Tkatchenko and M.~Scheffler, {\it Accurate molecular van der waals
  interactions from ground-state electron density and free-atom reference
  data},  {\em Phys. Rev. Lett.} {\bf 102} (2009), no.~7 073005.

\bibitem{zrte5}
T.~Matkovic and P.~Matkovic, {\it Constitutional study of the titanium,
  zirconium and hafnium tellurides},  {\em Metalurgija} {\bf 31} (1992),
  no.~107 110.

\bibitem{velden2004kenntnis}
A.~Velden and M.~Jansen, {\it Zur kenntnis der inversen perowskite m3to (m= ca,
  sr, yb; t= si, ge, sn, pb)},  {\em Zeitschrift f{\"u}r anorganische und
  allgemeine Chemie} {\bf 630} (2004), no.~2 234--238.

\bibitem{BNQ}
F.~Oton, R.~Pfattner, E.~Pavlica, Y.~Olivier, G.~Bratina, et~al.,
  \href{http://dx.doi.org/10.1039/C1CE05559C}{{\it Electronic and structural
  characterisation of a tetrathiafulvalene compound as a potential candidate
  for ambipolar transport properties}, } {\em CrystEngComm} {\bf 13} (2011)
  6597--6600.

\bibitem{monkhorst1976special}
H.~J. Monkhorst and J.~D. Pack, {\it Special points for brillouin-zone
  integrations},  {\em Phys. Rev. B} {\bf 13} (1976), no.~12 5188.

\bibitem{perdew1996}
J.~P. Perdew, K.~Burke, and M.~Ernzerhof,
  \href{https://link.aps.org/doi/10.1103/PhysRevLett.77.3865}{{\it Generalized
  gradient approximation made simple}, } {\em Phys. Rev. Lett.} {\bf 77} (Oct,
  1996) 3865--3868.

\bibitem{zheng2016}
G.~Zheng, J.~Lu, X.~Zhu, W.~Ning, Y.~Han, et~al.,
  \href{https://link.aps.org/doi/10.1103/PhysRevB.93.115414}{{\it Transport
  evidence for the three-dimensional dirac semimetal phase in
  $\mathrm{ZrT}{\mathrm{e}}_{5}$}, } {\em Phys. Rev.} {\bf B93} (Mar, 2016)
  115414.

\bibitem{Kotov}
V.~N. Kotov, B.~Uchoa, V.~M. Pereira, F.~Guinea, and A.~H. Castro~Neto,
  \href{http://dx.doi.org/10.1103/RevModPhys.84.1067}{{\it Electron-electron
  interactions in graphene: Current status and perspectives}, } {\em Rev. Mod.
  Phys.} {\bf 84} (Jul, 2012) 1067--1125.

\bibitem{gtpack1}
R.~M. Geilhufe and W.~Hergert,
  \href{https://www.frontiersin.org/article/10.3389/fphy.2018.00086}{{\it
  {GTPack: A Mathematica Group Theory Package for Application in Solid-State
  Physics and Photonics}}, } {\em Frontiers in Physics} {\bf 6} (2018) 86.

\bibitem{gtpack2}
W.~Hergert and R.~M. Geilhufe, {\em {Group Theory in Solid State Physics and
  Photonics: Problem Solving with Mathematica}}.
\newblock Wiley-VCH, 2018.
\newblock \textsc{isbn:} 978-3-527-41133-7.

\bibitem{Lomb:1976wy}
N.~R. Lomb, \href{http://dx.doi.org/10.1007/BF00648343}{{\it {Least - squares
  frequency analysis of unequally spaced data}}, } {\em Astrophys. Space Sci.}
  {\bf 39} (1976) 447--462.

\bibitem{Scargle:1982bw}
J.~D. Scargle, \href{http://dx.doi.org/10.1086/160554}{{\it {Studies in
  astronomical time series analysis. 2. Statistical aspects of spectral
  analysis of unevenly spaced data}}, } {\em Astrophys. J.} {\bf 263} (1982)
  835--853.

\bibitem{Fox:2011px}
P.~J. Fox, J.~Kopp, M.~Lisanti, and N.~Weiner,
  \href{http://dx.doi.org/10.1103/PhysRevD.85.036008}{{\it {A CoGeNT Modulation
  Analysis}}, } {\em Phys. Rev.} {\bf D85} (2012) 036008,
  [\href{http://arxiv.org/abs/1107.0717}{{\tt 1107.0717}}].

\bibitem{Heeba:2019jho}
S.~Heeba and F.~Kahlhoefer, {\it {Probing the freeze-in mechanism in dark
  matter models with $U(1)^\prime$ gauge extensions}},
  \href{http://arxiv.org/abs/1908.09834}{{\tt 1908.09834}}.

\bibitem{Vogel:2013raa}
H.~Vogel and J.~Redondo,
  \href{http://dx.doi.org/10.1088/1475-7516/2014/02/029}{{\it {Dark Radiation
  constraints on minicharged particles in models with a hidden photon}}, } {\em
  JCAP} {\bf 1402} (2014) 029, [\href{http://arxiv.org/abs/1311.2600}{{\tt
  1311.2600}}].

\bibitem{Vinyoles:2015aba}
N.~Vinyoles, A.~Serenelli, F.~L. Villante, S.~Basu, J.~Redondo, et~al.,
  \href{http://dx.doi.org/10.1088/1475-7516/2015/10/015}{{\it {New axion and
  hidden photon constraints from a solar data global fit}}, } {\em JCAP} {\bf
  1510} (2015), no.~10 015, [\href{http://arxiv.org/abs/1501.01639}{{\tt
  1501.01639}}].

\bibitem{Chang:2018rso}
J.~H. Chang, R.~Essig, and S.~D. McDermott,
  \href{http://dx.doi.org/10.1007/JHEP09(2018)051}{{\it {Supernova 1987A
  Constraints on Sub-GeV Dark Sectors, Millicharged Particles, the QCD Axion,
  and an Axion-like Particle}}, } {\em JHEP} {\bf 09} (2018) 051,
  [\href{http://arxiv.org/abs/1803.00993}{{\tt 1803.00993}}].

\bibitem{Gherghetta:2019coi}
T.~Gherghetta, J.~Kersten, K.~Olive, and M.~Pospelov, {\it {The Price of Tiny
  Kinetic Mixing}},  \href{http://arxiv.org/abs/1909.00696}{{\tt 1909.00696}}.
  
\bibitem{Chang:2019xva}
J.~H. Chang, R.~Essig, and A.~Reinert, {\it {Light(ly)-coupled Dark Matter in the keV Range: Freeze-In and Constraints}},  \href{http://arxiv.org/abs/1911.03389}{{\tt 1911.03389}}.

\end{thebibliography}
\end{document}